\documentclass[sigconf]{acmart}

\usepackage{tikz}
\usepackage{amsmath}
\usepackage[linesnumbered,ruled,vlined]{algorithm2e}
\usepackage{xspace}
\usepackage{multirow}
\usepackage{wrapfig}
\usepackage{booktabs}
\usepackage[skip=6pt]{caption}
\usepackage{amsthm}
\usepackage[inline]{enumitem}
\usepackage{url}

\usepackage{breakurl}
\usepackage{subfig}

\newenvironment{denseenum}{
\begin{enumerate}[topsep=2pt, partopsep=0pt, leftmargin=1.5em]
  \setlength{\itemsep}{2pt}
  \setlength{\parskip}{0pt}
  \setlength{\parsep}{0pt}
}{\end{enumerate}}

\newcommand{\paragrapha}[1]{\vspace{0.05in}\noindent{\bf #1}}

\newcommand{\todo}[1]{\textcolor{red}{#1}}

\makeatletter
\def\BState{\State\hskip-\ALG@thistlm}

\DeclareRobustCommand\onedot{\futurelet\@let@token\@onedot}
\def\@onedot{\ifx\@let@token.\else.\null\fi\xspace}
\def\eg{{e.g.,\xspace} \def\Eg{E.g.,\xspace}}
\def\ie{{i.e.,\xspace} \def\Ie{I.e.,\xspace}}

\makeatother
\newcommand{\sys}{PrismDB\xspace}
\newcommand{\name}{PrismDB\xspace}
\newcommand{\clock}{clock\xspace}

\newcommand{\Paragraph}[1]{\paragraph{#1.}}
\AtBeginDocument{%
  \providecommand\BibTeX{{%
    \normalfont B\kern-0.5em{\scshape i\kern-0.25em b}\kern-0.8em\TeX}}}

\setcopyright{acmcopyright}
\copyrightyear{2023}
\acmYear{2023}
\acmDOI{XXXXXXX.XXXXXXX}

\acmConference[Under Submission]{}
\acmPrice{2022}
\acmISBN{978-1-4503-XXXX-X/18/06}

\begin{document}

\title{Efficient Compactions Between Storage Tiers with \sys}

\author{Ashwini Raina}
\affiliation{%
  \institution{Princeton University}
  \country{}
}
\email{araina@cs.princeton.edu}

\author{Jianan Lu}
\affiliation{%
  \institution{Princeton University}
  \country{}
}
\email{jiananl@princeton.edu}

\author{Asaf Cidon}
\affiliation{%
  \institution{Columbia University}
  \country{}
}
\email{asaf.cidon@columbia.edu}

\author{Michael J. Freedman}
\affiliation{%
  \institution{Princeton University}
  \country{}
}
\email{mfreed@cs.princeton.edu}

\begin{abstract}
In recent years, emerging storage hardware technologies have focused on 
divergent goals: better performance or lower
cost-per-bit. Correspondingly, data systems that employ
these technologies are typically optimized either to be fast (but expensive)
or cheap (but slow). We take a different approach: by
architecting a storage engine to natively utilize
two tiers of
fast and low-cost storage technologies, we can
achieve a Pareto-efficient balance between performance and cost-per-bit.

This paper presents the design and implementation of \name, a novel key-value store that exploits two extreme ends of the spectrum of
modern NVMe storage technologies (3D XPoint and QLC NAND) simultaneously. 
Our key contribution is how to efficiently migrate and compact data between two different storage tiers. Inspired by the classic cost-benefit analysis of log cleaning, we develop a new algorithm for multi-tiered storage compaction that balances the benefit of reclaiming space for hot objects in fast storage with the cost of compaction I/O in slow storage.
%
Compared to the standard use of RocksDB on flash in datacenters today, \name's average throughput on tiered storage is 3.3$\times$ faster and its read tail latency is 2$\times$ better, using equivalently-priced hardware.


\end{abstract}

\ccsdesc[500]{Information systems~Storage architectures}
\ccsdesc[500]{Information systems~Data management systems}

\keywords{key-value stores, emerging storage}

\maketitle

\section{Introduction}
\label{sec:intro}


Several new NVMe storage technologies have recently emerged,
expressing the competing goals of improving performance and reducing storage costs. 
On one side, high performance non-volatile memory (NVM) technologies, such as Optane SSD~\cite{optane,unwritten-contract} and Z-NAND~\cite{znand}, provide single-digit $\mu$s latencies.
On the other end of the spectrum, cheap and dense storage such as QLC NAND~\cite{qlc,micronqlc} enables applications to store vast amounts of data on flash at a low cost-per-bit. Yet with this lower cost, QLC has a higher latency and is less reliable than less dense flash technology (\eg TLC NAND). 





Table~\ref{tab:cost-latency} compares the large variation in endurance, cost and performance across two representative storage technologies. 
For example, we observe that there is a roughly 65$\times$ performance difference between Optane SSD (NVM) and QLC on random reads, and sequential reads show a similar trend (not shown). However, Optane SSD costs more than 20$\times$ per GB compared to QLC. Endurance also varies widely: QLC NAND can only sustain a relatively small number of writes before exhibiting errors~\cite{micronwhitepaper}. 


%
Many studies have shown that simply running existing software systems on new hardware storage technologies often leads to poor results~\cite{MyNVM,flashield,KVell,splinterdb,spandb}. 
Therefore, significant recent effort has sought to build new software storage systems that are architected specifically for these new technologies~\cite{DIRECT,flashield,MyNVM,KVell,splinterdb,RIPQ}. 
They typically choose one point in the design space: fast but expensive~\cite{KVell,MyNVM,splinterdb} (\eg using Optane SSD or Z-NAND), or cheap but slower~\cite{DIRECT,flashield,RIPQ} (\eg using dense flash). However, these systems do not exploit the cost-performance benefits of using multiple storage tiers.

While a few recent key-value stores~\cite{mutant,spandb,exploitingoptane,orthus} combine fast and cheap storage technologies together, they reuse the same monolithic data structure used for flash (typically log-structured merge trees) and use naive techniques for compacting data across tiers (\eg by simply copying entire files from one tier to the other). 
As we demonstrate experimentally in \S\ref{sec:perf}, these existing approaches are ill-equipped for a multi-tiered use case, because they inefficiently use fast NVM devices.
In fact, we spent the first year of this project trying to retrofit a log-structured merge (LSM) tree to use multiple storage tiers, and take advantage of fast NVM devices by retaining more frequently-accessed data on NVM.
However, we were not able to show any performance improvement because doing so led to excessive compactions, hurting overall performance.

\begin{table}
    \centering
    \footnotesize 
    \begin{tabular}{l r r r r}
      \toprule 
      & NVM & QLC \\
      \midrule
      Lifetime (DWPD) & 200  & 0.1 \\
      Cost (\$/GB) & \$2.5  & \$0.1 \\
      Avg Read Latency (4KB) & 6$\mu$s & 391$\mu$s \\
      \bottomrule
    \end{tabular}
    \vspace{2mm}
    \caption{Comparing NVM (Optane SSD) and dense flash (QLC).
    Cost taken from cdw.com for Intel's Optane SSD 5800x and Intel 660p, and lifetime is based on publicly available information~\cite{660p,p5800x}. Latency of 4KB random read is computed with Fio~\cite{fio}.} 
    \label{tab:cost-latency}
    \vspace{-3ex}
  \end{table}

This led us to fundamentally rethink key-value store data structures and compaction mechanisms to fully exploit fast and slow storage tiers, in order to realize more optimal trade-offs between performance, endurance, and cost.
We design a new key-value store, \name, which 
assumes a setting where a large percentage (\eg 75-95\%) of the persistent storage capacity sits on dense flash (\eg QLC), and the rest on NVM (\eg Optane SSD). 
In such a setting, given the far inferior performance and endurance of flash compared to NVM, \name's primary design goals are to maximize read and write performance, while minimizing the amount of I/O (and especially writes) issued to flash.

To achieve this goal, \name uses a hybrid data layout, where hot objects are stored in slab-based files on NVM (where random access is fast), and cold objects are stored in a sorted log on flash (where sequential writes are prioritized). Since NVM does not suffer from write amplification and supports fast random writes, it can efficiently perform in-place updates and fresh inserts of objects directly into slabs. 
\name tries to store frequently read or updated data in NVM, and cold, immutable data on flash. To estimate the access frequency of objects, \name uses a lightweight object popularity mechanism based on the \clock algorithm. Yet since key popularity distributions vary across workloads, \name 
records \clock value distributions and uses them to determine which objects are colder and should be compacted to QLC.
As most requests are served from either DRAM or NVM, the bottleneck shifts from I/O to CPU. \name employs a partitioned, shared-nothing architecture to minimize the amount of synchronization between threads.



Since space in the faster tier (NVM) quickly fills up, \name needs to decide which objects to migrate from NVM to flash. This leads to a fundamental trade-off: keeping a large number of popular objects in NVM ensures a higher proportion of accesses are served from NVM, but it comes at the expense of migration or compaction efficiency. Retaining more objects in NVM means the system needs to work harder to find less-popular objects, and thus has to merge with a wider range of keys in flash, thereby increasing expensive flash write I/O. In addition, when deciding which keys to compact, \name needs to consider other typical factors~\cite{o1996log,LFS} that impact compaction performance: the overlap between the merged key ranges and amount of stale data to be cleaned. 

Our primary contribution is multi-tiered storage compaction (MSC), a novel compaction mechanism that 
captures the relationship between key popularity and write amplification on multi-tiered storage.
We design a new metric by adapting the classic cost-benefit model of the log-structured file system (LFS)~\cite{LFS} to a multi-tiered setting and assign a score to each range of keys that measures whether they are good candidates for compaction.
The score is higher when ranges have more cold data that can be compacted from the faster storage tier, and is lower when it incurs a large amount of flash I/O per compacted object. 
Since computing this metric precisely is expensive, \name introduces an approximation algorithm that performs well in practice (\S\ref{sec:msc}).
\name does not only migrate data from the faster tier to the lower tier, but also, under read-heavy workloads, it can promote objects from the slower tier to the faster one in response to changing object popularity distribution.

We implement \name and compare it to a multi-tiered version of RocksDB~\cite{rocksdb}, a version of RocksDB that uses NVM as a second-level (L2) cache, a version of RocksDB that pins hot objects to levels mapped on NVM, as well as to two academic systems - Mutant~\cite{mutant} and SpanDB~\cite{spandb}.
\name significantly outperforms all of its baselines under all YCSB~\cite{ycsb} workloads that issue point queries (puts/gets/updates) and under Twitter production traces~\cite{twitter-cache} that are insert-heavy or read-heavy.

Our paper makes the following primary contributions: 
\begin{denseenum}
    \item \textbf{Architecture.} A hybrid architecture that exploits NVM in a multi-tiered storage setup while minimizing writes to flash.
    \item \textbf{Multi-tiered compactions.} A model and algorithm for efficient compactions for multi-tiered storage that balances data placement with compaction I/O in the slow tier.
    \item \textbf{Popularity scoring.} A novel algorithm for popularity-based object placement decisions on multi-tiered storage.
    \item \textbf{Performance and cost-efficiency.} Compared to multi-tiered RocksDB, \name's more efficient data structures and compaction mechanism improve the throughput and average latency by 2.4$\times$ and 2.7$\times$ on write-dominated workloads, and by 2.5$\times$ and 2$\times$ on read-heavy workloads. On fsync enabled workloads, \name achieves 3$\times$ higher throughput than SpanDB.
\end{denseenum}


\section{Background and Directly Related Work}
\label{sec:bg}

We provide a brief background on new storage technologies, and survey systems that use multiple NVMe storage tiers. 

\paragrapha{Trends in storage.}
In recent years, NVMe storage devices have evolved in two orthogonal directions: faster (and more expensive) non-volatile memory and cheaper (and slower) dense flash. New fast NVM technologies, such as 3D XPoint~\cite{optane,micronqlc} and Z-NAND~\cite{znand}, which we refer to collectively throughout the paper as Non-Volatile Memory (NVM), provide random read and write latencies of 10$\mu$s or less. 

On the other end of the spectrum, NAND or flash technology has become ever more dense and cheap. Flash manufacturers have been able to pack more bits in each device, both by stacking cells vertically (3D flash), and by packing more bits per memory cell. However, making devices denser also causes their latency to increase and makes them less reliable~\cite{DIRECT,bleak,flashmemorysummit,micronqlc,660p,jaffer2020rethinking}. The latest QLC NAND technology, which packs 4 bits per memory cell, can only tolerate 100--300 write cycles before it becomes unusable~\cite{lowpe,flashmemorysummit,660p}. 
Future dense flash technologies, such as the recently announced PLC NAND (5 bits per cell), will exacerbate this trade-off~\cite{jaffer2020rethinking}.

\begin{figure}
  \begin{center}
  {\includegraphics[scale=0.2]{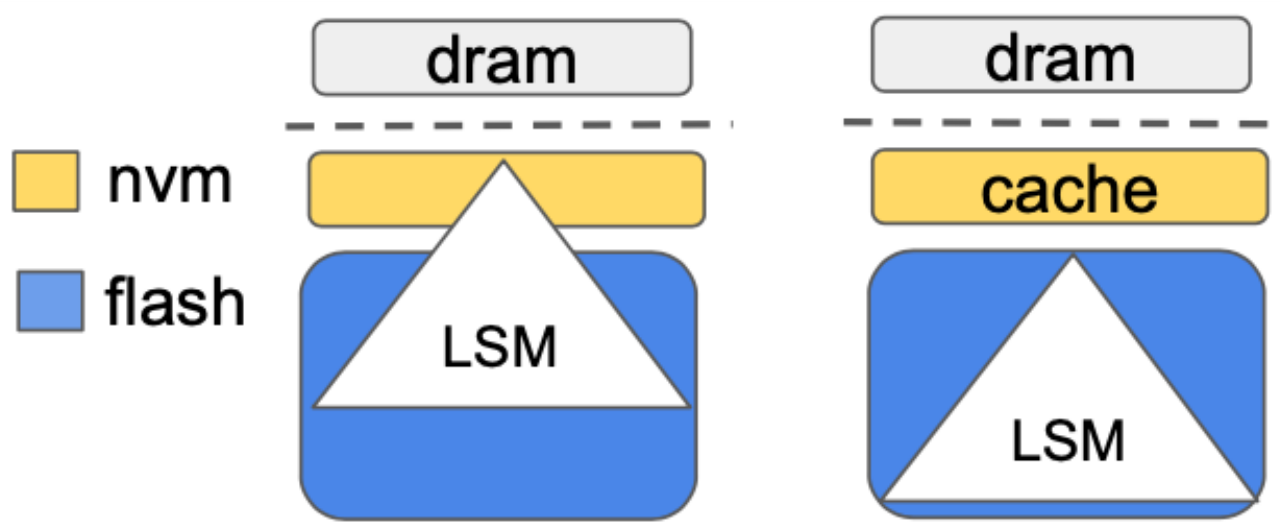}}
  \end{center}
  \caption{Tiered storage designs: embedded within a single-tier data structure (left) and extra cache capacity (right).}
  \vspace{-3ex}
  \label{fig:two-systems}
\end{figure}

\paragrapha{Existing tiered storage designs.}
With the emergence of fast NVMe storage devices, several recent storage system architectures were proposed for tiered storage. These designs typically try to augment an existing key-value store, which was typically originally designed for flash, with a small amount of fast storage (\eg Optane SSD).
We can broadly group the architectures into two categories, depicted by Figure~\ref{fig:two-systems}: 

\begin{denseenum}
\item \textbf{Embedded in data structure.} NVM is incoporated into an existing flash-based data structure. This approach is used by Mutant~\cite{mutant} and SpanDB~\cite{spandb}, both of which employ log-structured merge (LSM) trees. Mutant migrates cold LSM files to slow storage, while SpanDB stores the tree's top levels on NVM, and the bottom ones on flash.
\item \textbf{Extra cache.} NVM is simply treated as extra cache space for objects that are stored in DRAM. Example of such systems are: MyNVM~\cite{MyNVM}, SQL Server~\cite{exploitingoptane} and Orthus-KV~\cite{orthus}. MyNVM and SQL Server simply treat NVM as an L2 cache, while Orthus-KV dynamically uses NVM as an L2 cache or as auxiliary  storage tier that provides extra storage bandwidth.  
\end{denseenum}

In both of these approaches, a traditional single-tier data structure, which is optimized for flash (typically an LSM-tree), is retrofitted to use NVM. While these existing approaches have the advantage of relatively easy integration with existing flash-based key-value stores, they do not take full advantage of NVM. In the next section, we analyze why.

\section{LSM Performance on Emerging Storage}
\label{sec:perf}

Before describing why existing multi-tiered designs fail to achieve the full potential of NVM, we outline the desired properties of such a system.
For a persistent key-value store to be high-performance and affordable, it needs to satisfy the following design goals.

\begin{denseenum}
    \item \textbf{Navigate cost-performance trade-off.} The system should provide high performance (throughput, read/write latency), while most of its data (\ie 75-95\%) is on low-cost storage (\eg flash). 
    \item \textbf{Support small objects.} The system needs to support datacenter key-value workloads that consist of small objects (\ie 1~KB or smaller~\cite{fb2020,twitter-cache}). Thus, we cannot assume the database's index entirely fits in DRAM~\cite{splinterdb,MyNVM}. 
    \item \textbf{Minimize flash I/O.} Since flash is slow, the system should minimize flash reads. It should also reduce flash writes to maximize system lifetime given the limited write cycles on flash.
\end{denseenum}

We now evaluate these design goals on an LSM-based key-value store. We use RocksDB, a popular open-source key-value store~\cite{fb2020}. Throughout the paper, by default our experiments use YCSB-A~\cite{ycsb} (50\% reads, 50\% updates), on a 100~M key dataset, with 1~KB object sizes on the hardware setup described in \S\ref{sec:eval}.
\begin{table}
    \centering
    \footnotesize 
    \begin{tabular}{l | ccc | c }
      &  & RocksDB & & PrismDB \\
      & NVM & QLC & het & het\\
      \midrule
      Throughput (Kops/sec) & 121  & 54 & 93 & 184\\
      Cost (\$/GB) & \$2.5 & \$0.1 & \$0.3 & \$0.3 \\
      \bottomrule
    \end{tabular}
    \vspace{1mm}
    \caption{Comparing single-tier NVM and QLC with multi-tier (89\% QLC, 11\% NVM, labeled ``het'') on Zipf 0.8 workload.}
    \vspace{-3ex}
    \label{tab:lsm_setup_summary}
\end{table}





\paragrapha{Single-tier storage.}
We first analyze how a single-tier storage setup performs using NVM and QLC, the two extremes of the cost-performance trade-off. 
Table~\ref{tab:lsm_setup_summary} compares the throughput and cost of running RocksDB on the two storage configurations. We note that even though raw NVM performance is far superior to QLC (\eg 65$\times$ lower average 4~KB random read latency), the observed overall system throughput in Table~\ref{tab:lsm_setup_summary} is only 2.2$\times$ better than QLC. On QLC, RocksDB is I/O bound, and spends 36\% of its CPU time in iowait because QLC I/O is slow. On NVM, RocksDB is bottlenecked by CPU, since the system spends less time waiting for I/O completions. This leads us to ask: would a multi-tiered storage setup be bottlenecked by QLC I/O, CPU contention or both?

\begin{figure}[!t]
	\centering
	\subfloat[]{
	\label{fig:lsm-comp}
		\includegraphics[width=0.42\columnwidth]{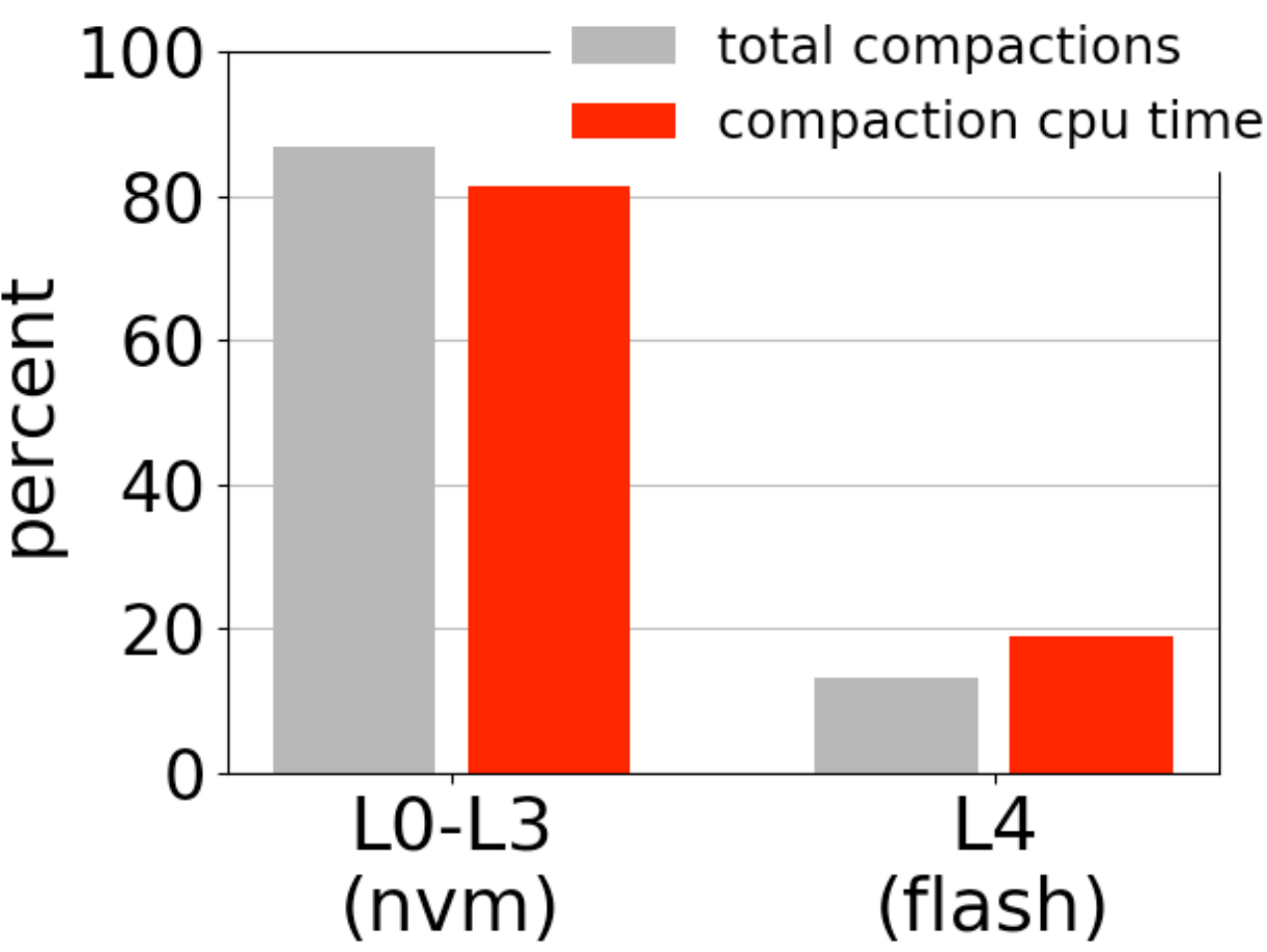}
	}~
	\subfloat[]{
	\label{fig:ycsb-reads}
		\includegraphics[width=0.41\columnwidth]{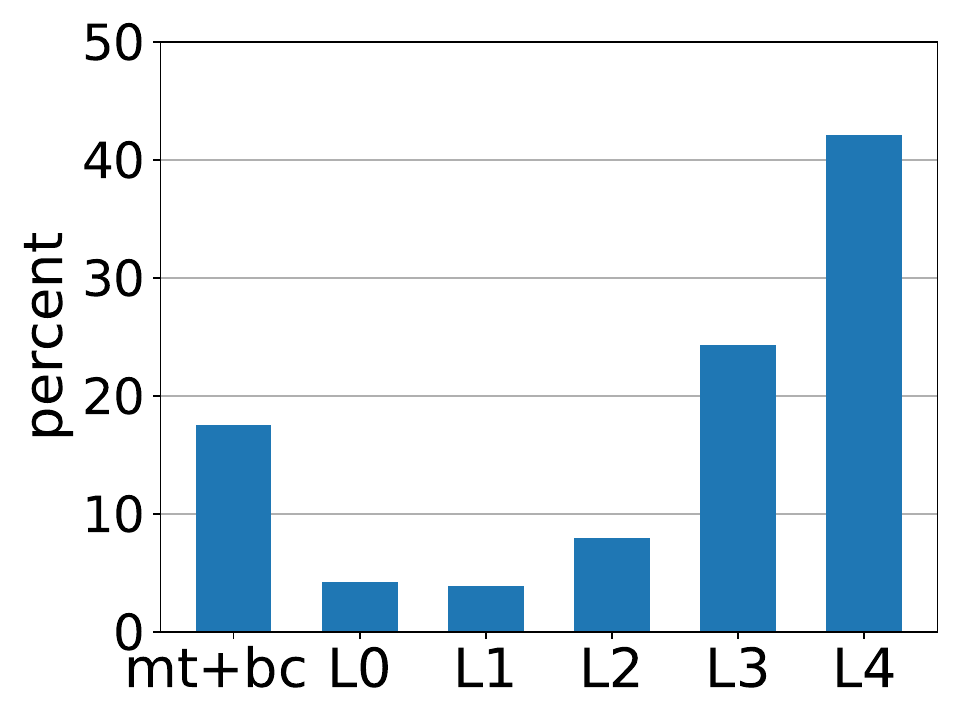}
	}
    \vspace{-1ex}
	\caption{(a) Percentage of compactions in NVM and QLC in a multi-tier RocksDB. (b) Distribution of reads across LSM levels (L0-L4), memtable (mt) and block cache (bc).}
	\vspace{-3ex}
	\label{fig:het}
\end{figure}


\paragrapha{Multi-tier storage.} 
We next examine how a LSM-tree performs on a multi-tiered setup, similar to the one used by SpanDB~\cite{spandb}.
We refer to the LSM multi-tiered storage setup as \emph{het} (\ie heterogeneous). 
We use a 5-level LSM-tree, in which levels L0-L3 are mapped to NVM, and L4 to QLC, where L4 is 89\% of the storage capacity in the database.

By storing only 11\% of the database on NVM, \emph{het}'s cost-per-bit (\$0.34/GB) is close to standard single-tier TLC flash setups used today in datacenters (\$0.31/GB).
At the same time, it achieves a throughput that is only 23\% less than the throughput of RocksDB running only NVM. 
Therefore, a multi-tier RocksDB configuration pays a small extra cost for faster storage but achieves a significant performance boost.

Nevertheless, the performance of RocksDB in this configuration is still far from optimal. We make two observations. First, as shown by others~\cite{rocksdbcidr,triad}, RocksDB spends significant CPU time (54\%) on background compactions. 
However, in our experiments with RocksDB in the tiered setting (Figure~\ref{fig:lsm-comp}), more than 80\% of compaction time is spent sorting data in the NVM tier. Since NVM supports fast in-place updates, spending CPU cycles sorting objects is unnecessary.



Second, significant CPU time (23\%) is still spent on I/O wait. Our results show (Figure~\ref{fig:ycsb-reads}), that even though RocksDB uses NVM for the upper levels of its tree, many reads are still served from flash (42\%), which is 65$\times$ slower than NVM, contributing to high I/O wait time. 
This negates the full performance benefit of using NVM as a fast storage tier. 
The reason more reads are not served from NVM is that LSM-tree organization is purely write-driven; LSM-trees do not try to cache frequently-read items on the upper levels. 

To add read-awareness to the LSM-tree design, we spent the first full year of this project building a prototype system, Rocksdb-RA, that stores more frequently-read objects on upper levels of the tree (L0-L3) on NVM using \emph{pinned compactions}. Unlike traditional LSM compaction that compacts all the objects down to the lower level, RocksDB-RA ``pins'' some percentage of the popular objects, retaining them in the NVM tier.



However, due to limited capacity of a level, retaining objects results in lesser free space and triggers more compactions on that level. We found that even though Rocksdb-RA was able to serve 27\% more client reads from NVM compared to flash, the total number of compactions increased by 2.3$\times$, which resulted in an overall degradation of the system's throughput and latency compared to RocksDB.

This leads us to conclude there exists a fundamental tension between object pinning and compaction efficiency, which we explore in this paper. 
Unfortunately, the salient features of an LSM tree, including its multiple levels and the need to use large sorted files, make compactions particularly expensive. As object pinning leads to more compactions, it typically degrades the performance in LSM trees. 
This motivates our decision to architect a new data layout and compaction mechanism that is tailored to multi-tiered storage.

\section{\name's Data Layout}
\label{sec:design}

In this section, we introduce \name's system components.

\begin{figure}
  \begin{center}
  {\includegraphics[scale=0.50]{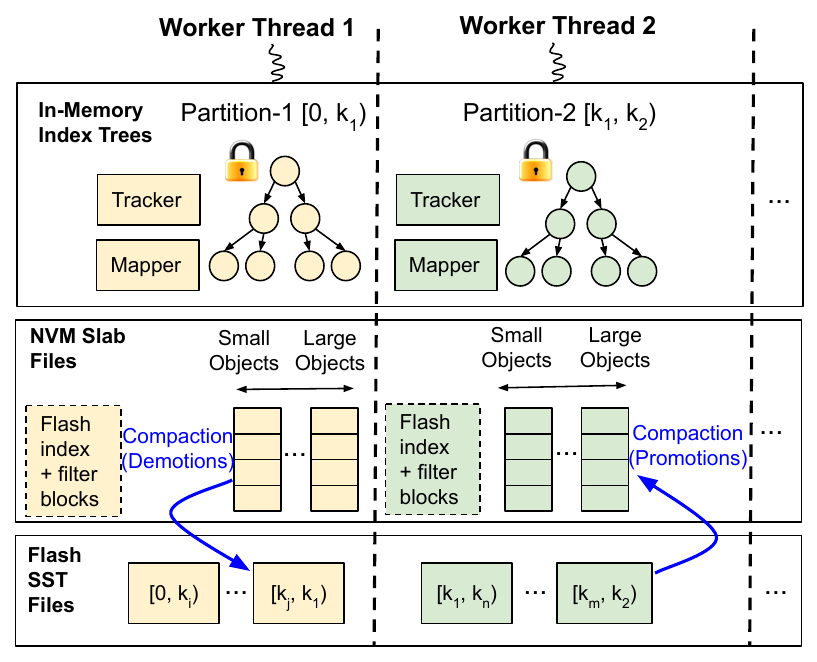}}
  \end{center}
  \vspace{-1ex}
  \caption{\label{fig:prismdb-sys} PrismDB system diagram.}
  \vspace{-3ex}
\end{figure}

\subsection{Design Overview} 
\label{sec:design-overview}

Figure~\ref{fig:prismdb-sys} depicts the architecture of \name on two-tiered storage (NVM and flash). 
Since NVM is low latency, in order to reduce synchronization on shared data structures, \name employs a partitioned, shared-nothing architecture.
Each partition consists of a subset of the key space and runs a dedicated worker thread to handle client requests one at a time.
This partitioned approach is extended across all storage tiers; \ie each partition handles its own data structures on DRAM, NVM, and flash. The partitioning algorithm can be range-based for scan heavy workloads or hash-based for workloads that exhibit load skew/imbalance.



\name uses a hybrid data layout, optimized for both tiers. 
At a high level, metadata (\eg indices and bloom filters) is stored on DRAM and NVM for fast lookups. 
DRAM and NVM also store recently-accessed key-value pairs. In order to support fast writes and reduce flash wear, all newly-written data (including updates) are written to NVM. Therefore, to optimize for fast random writes and in-place updates, the NVM data layout uses unsorted slabs. 
On the other hand, the data layout on flash
is optimized for sequential write access in order to minimize write amplification, and is therefore based on a sorted log. 

Each partition uses a \emph{tracker} and a \emph{mapper} that are stored in DRAM. The tracker estimates key popularity using the \clock algorithm. The mapper maintains the \clock value distributions and uses those distributions to enforce the placement of objects on storage. It uses a parameter, called the pinning threshold, to determine whether an object is ``popular'' enough for NVM. Each partition runs a background compaction thread to free up space on NVM by moving the colder objects to flash while retaining the popular ones on NVM.
Key ranges are selected for compaction based on a new compaction algorithm, called multi-tiered storage compaction (MSC), which we describe in \S\ref{sec:compactions}.
We now provide more detail on the data structures on DRAM, NVM, and flash.

\paragrapha{In-memory data structures.} We use a 
B-tree index to locate NVM objects, which are not sorted. Since we need to support a large database of small objects, an index over all keys in the database will not fit in memory. Hence, only the index for objects on NVM is stored in memory.
Each index entry in the B-tree stores the key and its NVM address (\ie 1-byte slab ID plus 4-byte page offset).
Each partition also stores the clock-based tracker in memory (see \S\ref{sec:tracker}).  
Each tracker entry includes the key and 1-byte \clock metadata.
\name does not use a userspace DRAM cache for caching frequently-read objects and instead relies on the OS page cache.

\paragrapha{Data layout on NVM.}
All new data is written to NVM, and it also serves as a second-level read cache after DRAM for hot objects.
NVM uses a slab-based data layout to support fast random inserts and updates of small objects.
\name uses a set of slab files each of which is dedicated to a specific object size range (\eg 100B, 200B, \ldots, 1KB).
Objects of similar size are inserted into available fixed-size slots in the right slab file, along with a metadata header containing version number (implemented as logical timestamp) and object size information.
If an object is deleted, its slot is freed for new data. 
An in-place update can take place on the original slot if the object doesn't change its size range, and if it does, the object needs to be deleted and moved to another slab file.


The index for objects on flash is stored in NVM rather than DRAM, because reading them from flash (\ie hundreds of microseconds) amortizes the cost of the index lookup. 
Similarly, \name stores on NVM a bloom filter~\cite{bloomfilter} for each file on flash to prevent issuing expensive flash I/Os for non-existent objects. The combined size of the flash index and filters can range from 100s of MBs to a few GBs, depending on the object size. For a small-object database, the flash index and filters reside entirely on NVM.
  
\paragrapha{Data layout on flash.} To support large sequential writes, data on flash uses Sorted String Table (SST) files (similar to LSM trees~\cite{leveldb,rocksdb}), which are stored in a log. 
Each SST file has an index that points to all the file's data blocks, and a bloom filter for the file's objects. SST files store disjoint key ranges, which makes searching an object fast.
When the percentage of NVM in \name is 10\% or higher, by default we store all the flash data in a single-level log.  When the percentage is lower, \name by can store the flash data in a multi-level log, similar to an LSM-tree. This choice is based on the fanout of objects stored on NVM and flash.  For example, a uniform workload with 10\% of the database in NVM, would yield key ranges in NVM that overlap with key ranges in flash that are 10$\times$ their size.  Therefore, lower NVM capacity will cause higher flash write amplification.

\subsection{Lifecycle of a Written Object}
\begin{figure}
  \begin{center}
    {\includegraphics[scale=0.2]{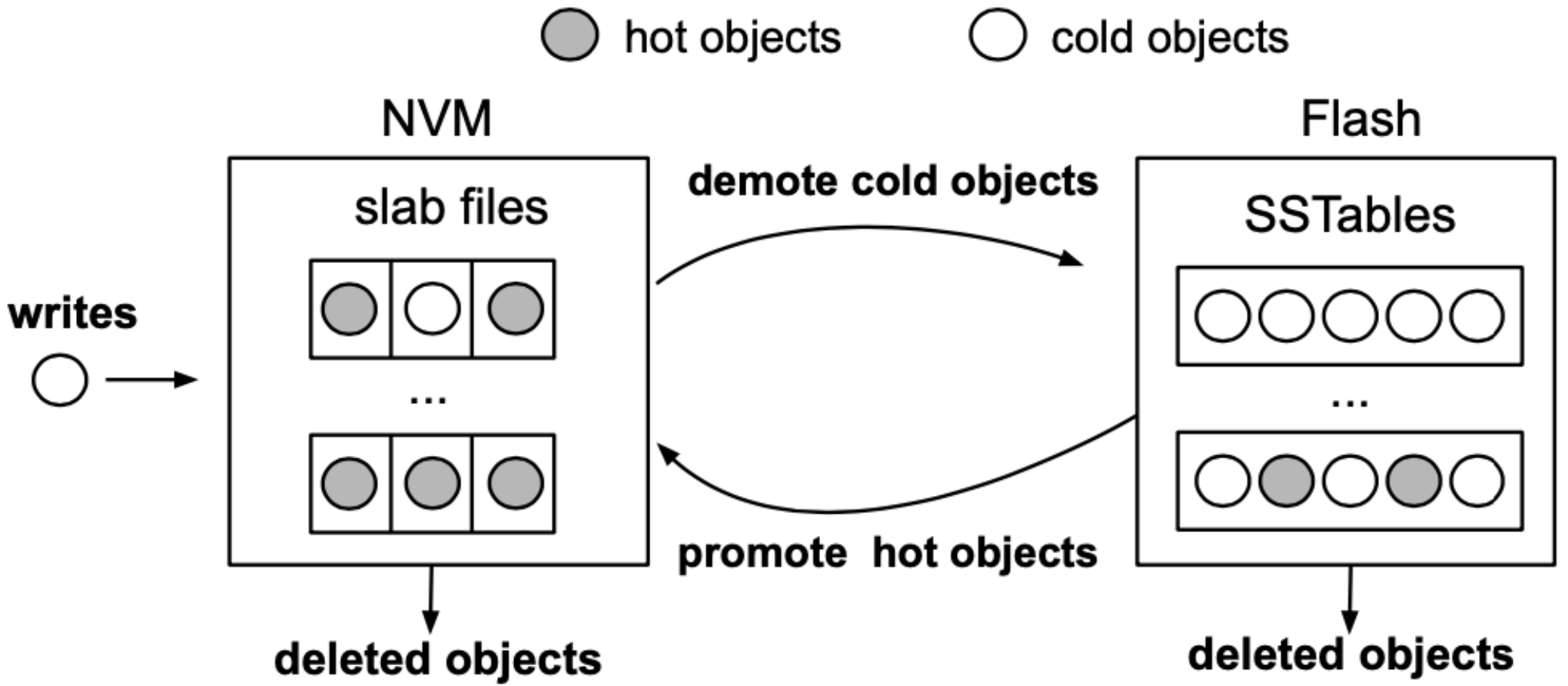}}
  \end{center}
  \vspace{-1ex}
  \caption{\label{fig:write-lifecycle} Lifecycle of a written object in \name.}
  \vspace{-3ex}
\end{figure}

Figure~\ref{fig:write-lifecycle} illustrates the lifecycle of an object written to \name. 
Objects are always synchronously written to NVM because writes need to be persisted for crash recovery. 
When NVM's used capacity hits a high watermark (by default 98\%), the partition triggers a background compaction job to demote colder objects to flash until it frees up enough space on NVM (by default when NVM usage reduces to 95\%).
In the meantime, incoming writes are rate-limited to ensure NVM does not exceed its capacity. 
The job selects ``cold'' NVM objects, by filtering the hot ones using the mapper, and \emph{demotes} them to flash.  
At the same time, given that \name is already incurring the cost of writing a new SST file, if the compaction job finds hot flash objects, it may \emph{promote} them to NVM during the compaction. In read-heavy workloads, where write-triggered compactions are rare, \name proactively triggers compactions when it detects that too many objects are accessed from flash, with the goal of promoting hot objects to NVM.
Since NVM stores more recent data, obsolete versions of objects on flash are deleted when merging with new versions from NVM. 

\subsection{Popularity Tracking} 
\label{sec:popularity-tracking}
We now discuss in more detail how \name tracks popular objects, and designates them as hot (or not).



\paragrapha{Tracker: lightweight tracking of objects.} 
\label{sec:tracker}
The tracker estimates object access popularity, while incurring a minimal overhead. There is a large body of work on how to track and estimate object popularity~\cite{LHD,RIPQ,mini-caches,SHARDS,lrb}.
However, many existing mechanisms require a relatively large amount of data per object, and computation per access.
Given that key-value objects are often small (\eg less than 1~KB~\cite{cliffhanger,nishtala2013scaling,fb2020}), we need to limit the amount of metadata we use for tracking purposes per object. We also need to be able to track millions of objects at a high throughput. 

We turn to \emph{\clock}~\cite{clock}, a classic approach to approximate the least recently used (LRU) eviction policy while offering better space efficiency and concurrency~\cite{fan2013memc3,flashield}.
\name's tracker uses the multi-bit \clock algorithm for object tracking.
The tracker uses a concurrent hash map that maps object keys to their \clock bits. The object values are not stored in the tracker.
Each client read or update operation requires the tracker to update the \clock bits of the object that was accessed. Once the tracker becomes full, it uses the \clock algorithm to decrement clock values of objects and then evicts the object with value 0.

Since setting the \clock bits is on the critical path of reads and updates, the tracker is optimized for concurrent key insertions, evictions, and lookups. Further, to save space, the tracker does not store \clock bits of all key-value pairs in the database, only the most recently accessed. In our evaluation, tracker size is set to 10\% of the total database keys; see \S\ref{sec:impl} for more implementation details.



\begin{figure}[t!]
    \begin{center}
    {\includegraphics[width=0.50\columnwidth]{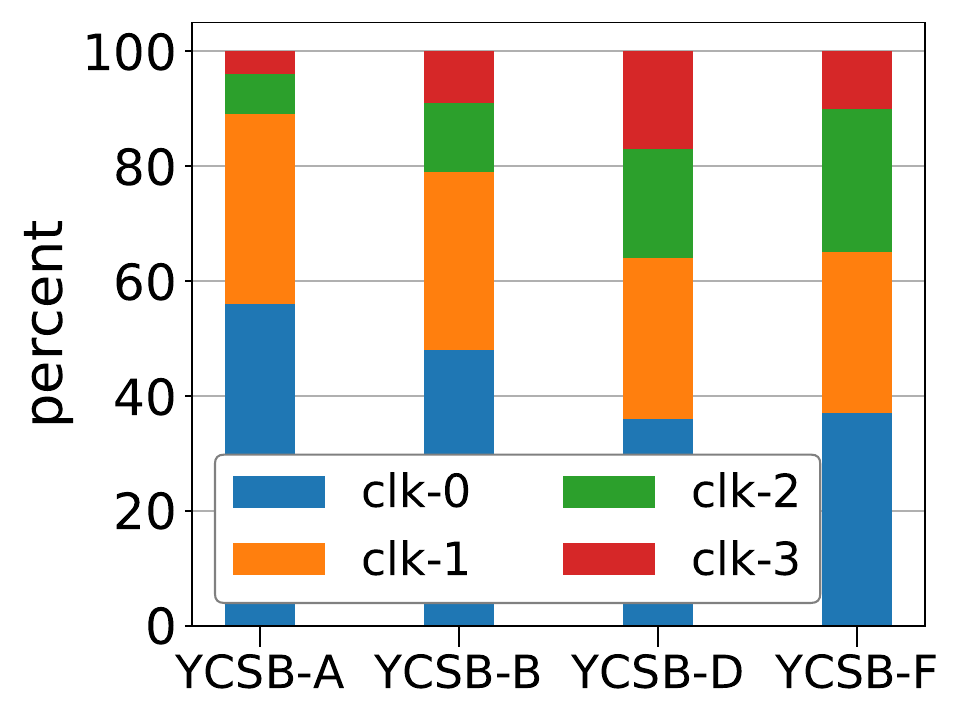}}
    \end{center}
    \vspace{-1ex}
    \caption{Clock value distributions under different workloads.}
    \vspace{-3ex}
    \label{fig:clock}
\end{figure}

\paragrapha{Mapper: enforcing the pinning threshold.}
Ideally, \name needs a threshold to determine which objects are hot, which we call the \emph{pinning threshold}.
In other words, at each pass of the compaction job, \name should pin on faster storage some percent (\eg 10\%) of the most popular objects that are being tracked.

However, enforcing this threshold depends on the \clock bit distribution, which vary as a function of the workload (Figure~\ref{fig:clock}).
Consider a two-bit \clock, with 3 being most popular and 0 being least popular. If \name wants to enforce a pinning threshold of 10\%, and exactly 10\% of keys have all their \clock bits set to 3, then \name should pin all the objects with a \clock value of 3. However, if 50\% of the keys have a \clock value of 3, then \name cannot pin all objects a value of 3, otherwise it will exceed the desired pinning threshold. 
To this end, the mapper is responsible for keeping track of the \clock value distribution, and uses that distribution to enforce the pinning threshold.

\paragrapha{Pinning threshold algorithm.}
In order to enforce the pinning threshold, the mapper uses the following algorithm, which is best illustrated with an example. 
Suppose the \clock distribution is similar to the YCSB-B workload depicted in Figure~\ref{fig:clock},
where the percentage of keys with a \clock value of 3 is about 10\%, the percentage of those with a value of 2 is 10\%, the ones with a value of 1 is 30\%, and the remaining 50\% have a \clock value of 0. 

Now, suppose the desired threshold is 15\% (\ie the most popular 15\% of objects are stored on NVM, and the rest on flash). If the compaction job encounters an object with a \clock value of 3, it will always pin it. If it encounters an object with a \clock value
of 2, it will randomly choose whether to keep it or not (in this example, with a probability of 0.5). If it encounters either an object with \clock value of 1 or 0, or an object that is not currently being tracked (recall the tracker does not track all objects in the database), that object will be demoted to flash.

To summarize, the mapper satisfies the pinning threshold using the highest-ranked \clock objects by descending rank, and if need be, randomly samples objects that belong to the lowest \clock value that is needed to satisfy the threshold.  

\section{Multi-tiered Storage Compaction}
\label{sec:compactions}

In this section we describe \name's compaction mechanism. We discuss the trade-off between garbage collection and data pinning (\S\ref{sec:trade-off}). We then present an analytical model for multi-tiered compaction efficiency (\S\ref{sec:compaction-efficiency}). Then we present our multi-tiered storage compaction algorithm (MSC) (\S\ref{sec:msc}).

\subsection{Performance Trade-off}
\label{sec:trade-off}
\name's compaction serves two purposes. 
First, it needs to reclaim enough space in NVM to absorb new incoming writes. Compaction frees up space in NVM when it is close to running out of capacity, and removes stale values from flash that have been updated more recently in NVM. Second, it needs to prioritize limited NVM space for hot data. Compaction demotes cold objects to flash and promotes hot objects to NVM.

However, these two functions present a fundamental trade-off.
On one hand, if \name were to move objects from NVM to flash without considering their popularity, then NVM will not serve as an effective second-level cache.
On the other hand, if \name were to pin a high percentage of objects that it encounters, it will take it much longer (and require more CPU) to free up enough space in NVM. This will not only delay incoming client writes, but also incur higher flash write amplification. This is because the objects in NVM selected to be demoted will have a ``sparser'' key range, and require the compaction job to merge the NVM objects with a larger number of SST files on flash.

\subsection{Modeling Compaction Cost and Benefit}
\label{sec:compaction-efficiency}

Since data on NVM is organized as individual key-value pairs and unpopular objects can be scattered across the entire key space, there are many possible choices of which objects to select for compaction. 
To reduce the search space and bound the flash I/O caused by each compaction, \name divides the contiguous NVM key space into smaller key ranges based on existing SST file bounds on flash. 
We define a compaction key range, $i$, as the key ranges of $i$ consecutive SST files. 
The value of $i$ is tunable in \name, and by default is set to 1.
A higher value of $i$ is more suitable for workloads with smaller SST file size or with many popular objects evenly distributed across the global key space.

\paragrapha{Analytical model.}
We design a novel compaction policy, which is inspired by the classic log cleaning cost-benefit analysis~\cite{LFS}.
In traditional log cleaning or compaction (\eg LFS~\cite{LFS} and other related systems~\cite{memshare,lsm-ramcloud}), the system tries to select the optimal contiguous segment of data to compact. It tries to choose the segment that offers the highest benefit (free up the most space on disk, and keep it free for a long time), at the lowest cost (I/O incurred by the garbage collection).
We adopt a similar approach, but we model the benefit and cost differently, adapting them to the multi-tiered setting.

\paragrapha{Benefit.}
We model the benefit as demoting as many ``cold'' objects to the slow storage tier as possible.
Cold objects are ones that have not recently been accessed either by a read or a write.
We consider writes, because objects that are not frequently updated are likely to stay ``stable'' in the future, which avoids moving them back and forth across tiers thereby saving costly flash disk bandwidth.
Thus, compaction offers greater benefit to the system if it can move more cold data to the slower tier.

We assign every object a \emph{coldness} score between (0, 1] (where 1 is cold and 0 is hot).
The coldness of object $j$ is the inverse of its clock value incremented by 1, $coldness(j)=\frac{1}{clock_{j} + 1}$. We increment the clock value by 1 to avoid dividing by zero. 
If the object doesn't appear in the tracker, we assume its clock value is 0 and its coldness score is therefore 1. 

We define the multi-tiered compaction \emph{benefit} of a key range as the the sum of coldness values of the objects in that range: \(\text{benefit} = \sum_{j=1}^{t_n}{coldness(j)}\), where $t_n$ is the number of objects in the key range. 
Table~\ref{tab:notation} summarizes the notations and their meanings.

\paragrapha{Cost.}
Log-structured systems like LFS typically consider the cleaning cost as the extra
I/O incurred on the same disk where space is freed. 
In our multi-tiered disk setup, however, I/O on the slower tier is an order of magnitude or more costlier (in terms of bandwidth, latency and endurance) than I/O on the faster tier. 
Thus, for simplicity, we model the cost of compaction as total flash I/O incurred per migrated byte from NVM. 
During compaction, older versions of objects on flash will first be read and later get deleted when merging with more recent NVM data.
Thus, compacting an NVM key range involves reading all SST file objects from flash, and then writing unpopular NVM objects and non-overlapping SST file objects back to flash.

We initially assume object sizes to be equal, and define notations in terms of ``number of objects'', which directly equates to their size. 
For an NVM key range where $p$ is the ratio of popular objects, the number of unpopular objects is \((1-p) \cdot t_{n}\), where $t_{n}$ is the number of objects in that candidate NVM key range.
We define $o$ as the fraction of objects in the SST file that also appear in the NVM key range. Then the number of non-overlapping objects in the SST file becomes \((1-o) \cdot t_{f}\), where $t_{f}$ is the total number of objects in the flash SST file. 
Thus compacting \((1-p) \cdot t_{n}\) objects on NVM incurs \(t_{f}\) read I/O from flash and \((1-p) \cdot t_{n} + (1-o) \cdot t_{f}\) write I/O to flash.
The flash I/O per migrated object is: \(\frac{t_{f} + (1-p) \cdot t_{n} + (1-o) \cdot t_{f}}{(1-p) \cdot t_{n}} = \frac{(2-o) \cdot t_{f}}{(1-p) \cdot t_{n}} + 1\)
We define \(\frac{t_{f}}{t_{n}}\) as the fanout, $F$, which represents the size ratio of key ranges on NVM and flash.
Therefore, the flash I/O cost is reduced to \(F \cdot \frac{ (2-o)}{(1-p)} + 1\).

Finally, we define our multi-tiered storage compaction (MSC) metric as the ratio of benefit to cost. 
\vspace{-1ex}
\begin{equation}
    \text{MSC} = \frac{\text{benefit}}{\text{cost}}
    = \frac{\sum_{j=1}^{t_{n}}{coldness(j)}}{F \cdot \frac{ (2-o)}{(1-p)} + 1} 
    \label{eq:migration}
\end{equation}
The metric's score is higher for key ranges that contain more objects that are colder and when it can incur lower I/O overhead on flash per migrated object.
Given a list of candidate compaction key ranges, \name selects the range with the highest score for compaction. 


For workloads with variable-sized objects, $p$, $o$ and $F$ are normalized by size (in bytes) to compute MSC precisely.


\begin{table}[t!]
    \centering
    \footnotesize
    \begin{tabular}{p{0.6cm} p{6.9cm} }
      \toprule 
       Symbol & Description\\
       \midrule
       $\; clock_{j}$ & Clock value of a NVM object $j$.\\
       $\; t_{n}$ & Total number of objects in selected NVM key range. \\
       $\; t_{f}$ & Total number of objects in SST file before merging.  \\
       $F$ & The fanout ratio $\frac{t_{f}}{t_{n}}$ between key range on flash and NVM.\\
       $p$ & Fraction of popular objects in candidate NVM key range.\\ 
       $o$ & Fraction of overlapping objects between NVM range and flash.\\
       \bottomrule
    \end{tabular}
    \caption{Notations.}
    \vspace{-1ex}
    \label{tab:notation}
\end{table}

\subsection{MSC Algorithm}
\label{sec:msc}

\begin{figure}[t!]
	\begin{center}
	{\includegraphics[width=0.9\columnwidth]{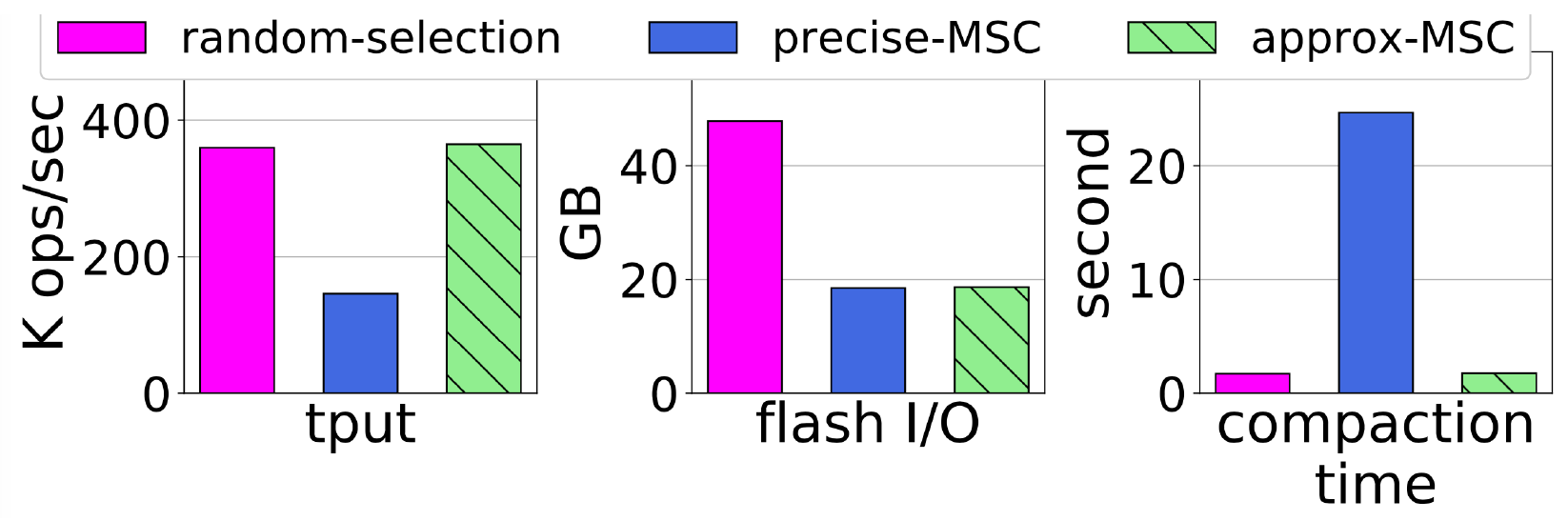}}
	\end{center}
	\vspace{-1ex}
	\caption{Comparison of throughput, flash write I/O and average compaction time between \name's precise-MSC and approx-MSC metric and the random-selection policy under YCSB-A with Zipf 0.99.}
	\vspace{-3ex}
	\label{fig:mig-metric}
\end{figure}

\paragrapha{Precisely computing the compaction metric.}
To test the MSC metric, we first implement a policy that precisely scores all candidate key ranges and selects the one with highest MSC score.
We also introduce a strawman policy, random-selection, that randomly selects a candidate key range and moves its cold objects to flash.
Figure \ref{fig:mig-metric} shows the comparison of precise metric with the strawman.

Since the random-selection policy is unaware of compaction benefit or cost, it can choose key ranges that contain fewer cold objects and incur higher write amplification on flash. 
We observe that the precise-MSC metric can decrease flash write I/O by more than 2.5$\times$ compared to the random-selection policy.
However, it has worse overall throughput.
This is because computing MSC precisely in Equation \ref{eq:migration} requires checking the popularity of each object in the mapper, and navigating the indices of all items in the candidate range both in the DRAM B-tree and in the SST file indices, to check for overlaps.
This is CPU intensive and leads to long compaction time (25 seconds, compared to the random-selection policy's 1.7 seconds), during which \name has to rate-limit foreground client writes until it frees enough space on NVM.

\paragrapha{Approximating compaction metric.}
Therefore, we propose a light-weight metric called approx-MSC that efficiently approximates the value of MSC.
Instead of computing the statistics for each individual object in the candidate key range, approx-MSC metric breaks each partition's key space into smaller, fixed-sized ranges, which we call buckets. 
It then keeps tracks of approximate values related to $p$, $o$, and $F$ for each bucket.
approx-MSC can be computed using a weighted sum of the parameters of each bucket that overlaps with the candidate key range. 
We describe the implementation and maintenance of buckets in \S\ref{sec:impl}.


Figure \ref{fig:mig-metric} shows that approx-MSC achieves high throughput while keeping total flash write I/O low (\ie nearly same as precise-MSC's). 
Since its approximation takes less time and fewer CPU resources, it reduces average compaction time from 25 seconds to 1.7 seconds, close to the random-selection policy. Therefore, we use approx-MSC in \name, and henceforth in the paper when we refer to MSC we are referring to approx-MSC.

\paragrapha{Key range selection.}
Another important design question is which NVM key ranges should be considered as candidates for compaction. 
One simple approach is to enumerate over all possible NVM key ranges to decide which to compact, but this is impractical for large databases. 
We use power-of-k choices~\cite{power-of-two} to select a subset of compaction key ranges as candidates. We empirically use $k=8$, as it provides a good trade-off between throughput and flash I/O. 
\paragrapha{Object promotions under dynamic workloads.}
During the compaction process, objects can be both promoted and demoted. Demotions have an obvious benefit: they free up space in NVM. Promotions, on the other hand, are more expensive, since they take up space in NVM for an object that was previously stored in flash.
However, sometimes objects also need to be promoted from flash to NVM, in order to enable fast reads for popular objects, which at some point got demoted. For example, this would occur when a large burst of newly-written objects fills up NVM. Such bursts have been observed in production workloads~\cite{fb2020,twitter-cache,cliffhanger}. 

\paragrapha{Read-triggered Compactions.}
By default, compactions are triggered when NVM fills up.
However, this is not always the best trigger for moving items between the two tiers. In read-heavy workloads, for examle, NVM will only slowly fill up, and the write-triggered compaction process may not be called frequently enough to keep up with changing read popularity distributions.
Therefore, \name also employs read-triggered compactions for moving items between NVM and flash. The main goal of read-triggered compactions is to improve the ratio of reads served from NVM. 

Read-triggered compactions have three stages: detection, invocation, and monitoring. In the detection stage, \name checks if the write-triggered compactions aren't keeping up with popularity distributions, by detecting read-dominated workloads where a large proportion of the keys in the tracker are stored on flash. If so, it triggers compactions for an epoch (by default 1~M client operations). At the end of each epoch, the monitoring stage tracks the ratio of reads served from NVM vs. flash. If the compactions that were triggered in the previous epoch improved that ratio above a threshold (by default 1\%), it continues the compactions for next epoch. If not, it enters a cool-down period (by default 10~M operations). At the end of that period, read-triggered compactions resume again from the detection stage. 

\section{Implementation}
\label{sec:impl}

\begin{figure}
  \begin{center}
    {\includegraphics[scale=0.3]{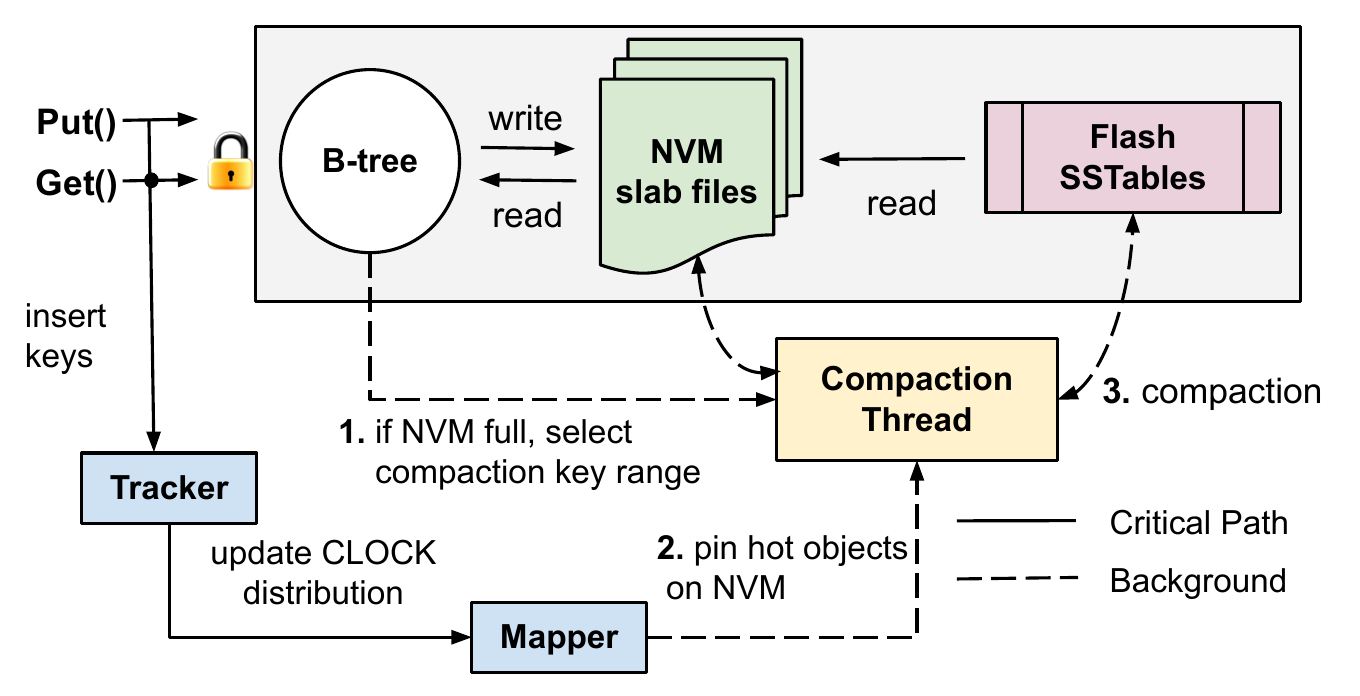}}

  \end{center}
  \vspace{-1ex}
  \caption{\label{fig:impl-overview} \name's system components.}
  \vspace{-3ex}
\end{figure}

In this section, we describe \name's implementation.
\name \footnote{https://github.com/princeton-sns/prismdb} is implemented 
in C++. It is built on top of Google's B-tree implementation~\cite{btree}, the SST file format from LevelDB~\cite{leveldb} and the slab implementation for NVM from KVell~\cite{KVell}. 
We depict the implementation's principal components in Figure~\ref{fig:impl-overview}.
Every partition in \name runs two threads, both of which synchronize using a single partition lock. 
A worker thread (depicted as a straight line in the figure) handles foreground client operations, including looking up, reading objects and writing data to NVM. 
The worker thread is also responsible for tracking object popularity and its distribution in the tracker and mapper. 
A compaction thread, depicted as a dashed line, runs in the background and is triggered intermittently to free space on NVM and re-balance data popularity so hot objects are always stored on NVM. 

\paragrapha{Interface.} \name supports 4 types of operations: \texttt{Put}, \texttt{Get}, \texttt{Scan} and \texttt{Delete}. 
Client requests are forwarded to the appropriate partition worker based on the operation key.
The worker thread always acquires the partition lock before processing the request and releases it at the end.

\texttt{Put(k,v)} writes the key $k$ with value $v$ to NVM.
The worker thread checks if the partition's B-tree index has that key.
If the key is not present, it selects the slab file based on the object size, and inserts the object to a free slot location within the slab. Then it inserts the key and its disk location in the partition's B-tree index. 
If the key is present, the worker thread checks if the new object size still fits within the original slab size range. 
If it does, it performs an in-place update to the same disk location. 
Otherwise, it deletes the object from the current slab and performs a fresh insert to the new slab. 
Then it updates the object's B-tree index entry with new disk location.
Lastly, the worker thread updates the key popularity in the tracker.


\texttt{Get(k)} returns the the most recent version of the key if found. The worker thread checks if the key is in the B-tree index.
It the key is present, it reads the object from the slab file on NVM.
Otherwise, it checks the index and filter blocks of the SST files and reads the object from flash. 
Lastly, the worker thread updates the key popularity in the tracker. 

\texttt{Scan(k,n)} fetches the next $n$ objects with keys equal to or greater than $k$. The worker thread runs a two-level iterator, one on NVM objects and the other on flash objects.  In every iteration, it compares the key values of objects pointed by the two iterators and selects the smaller object, and then moves the corresponding iterator pointer forward. For range queries that span multiple partitions, \name locks and scans one partition at a time. 

\texttt{Delete(k)} first looks up the key. If the key exists in NVM, it is deleted from the B-tree index and its slab slot is reclaimed. If the key is present on flash, the worker thread performs a fresh insert to NVM with a special delete tombstone entry. Eventually, both the NVM tombstone and the flash object will be deleted when compaction merges them.

\paragrapha{Tracker and mapper.}
The tracker is built on the concurrent hash map implementation from Intel's TBB library~\cite{tbb}. 
The hash-map index is the object key and each index stores a 1~byte value - two bits for clock and one bit for object location (NVM or flash). 
When a database client calls \texttt{Get} or \texttt{Put}, the worker thread inserts the key into the tracker. An insert operation can also invoke an eviction. Keys are initially inserted into the tracker with a value of 0 (min popularity), and keys accessed afterwards have their value set to 3 (max popularity). During eviction, clock values get decremented as per the \clock algorithm, and keys with clock values of 0 are evicted.

The \clock value is stored as an atomic variable. Thus, looking up a \clock value does not need to be serialized with eviction.
The mapper is implemented as an array of four atomic integers; each keeps track of the number of keys with a particular \clock value.

\paragrapha{Compaction thread.}
Once triggered, the compaction thread first acquires the partition lock. 
It uses MSC to select a compaction key range and only picks unpopular objects for compaction.
Next, the compaction thread reads these objects from NVM and the overlapped SST file(s) to memory, merge-sorts them and rewrites live objects as new SST file(s) back to flash.
This stage consumes the most time in the compaction (seconds). 
Therefore, the compaction thread releases the lock before merging the files. We use a reference counting scheme, similar to RocksDB~\cite{versionset} to track live SST files in flash. This guarantees that compaction doesn't delete a SST file that is being used by a concurrent \texttt{Get} or \texttt{Scan} iterator. 
%
Concurrent client writes can update a compacted object with a more recent version. The compaction thread ensures data correctness by re-acquiring the partition lock to check if the version on NVM has changed, and if it has, it skips deleting that item.
To do so, \name uses a lightweight compaction bitmap to track whether a object has changed its value since compaction. 

\begin{figure}
  \begin{center}
    {\includegraphics[scale=0.3]{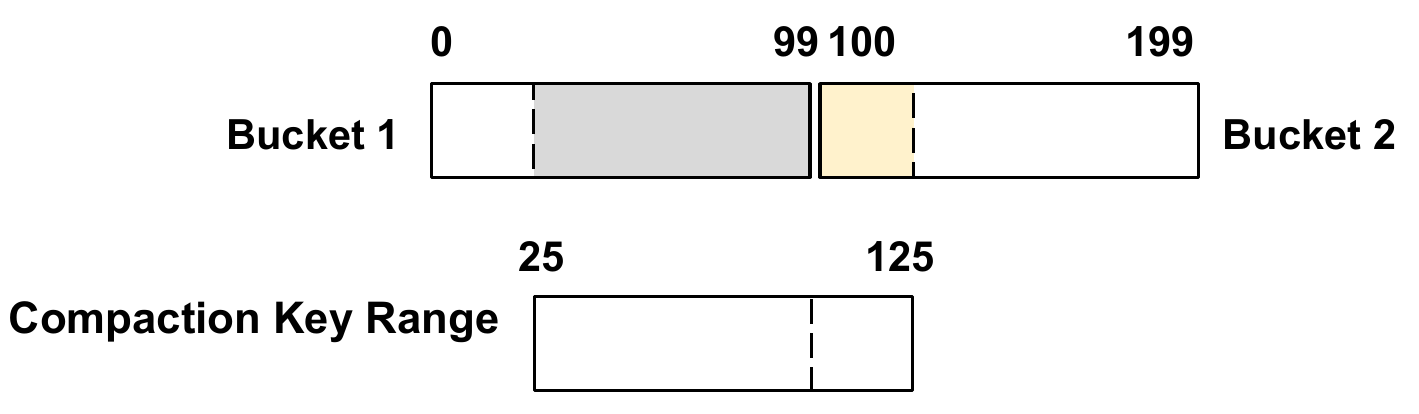}}
  \end{center}
  \vspace{-1ex}
  \caption{\label{fig:impl-msc} Buckets and compaction key ranges.}
  \vspace{-3ex}
\end{figure}

\paragrapha{MSC metric.}
By default, the approx-MSC metric uses a bucket size of 64K keys, which is equal to the average number of keys in an SST file. 
The global key space is divided into consecutive buckets. 
Figure~\ref{fig:impl-msc} shows a toy example with bucket size of 100 keys and a candidate compaction key range that overlaps with two buckets. 
Every bucket contains four fields, each with initial values of 0: \texttt{num\_nvm\_keys} (a counter of the number of keys present on NVM), \texttt{pop\_bitmap} (a bitmap of key popularity), \texttt{nvm\_bitmap} (a bitmap of keys on NVM), and \texttt{flash\_bitmap} (a bitmap of keys on flash). 
Access to these fields is protected by the partition lock.

\texttt{Put}s increment \texttt{num\_nvm\_keys} in the corresponding bucket.
Meanwhile, compactions decrement this parameter for each overlapped bucket since they know exactly which keys to remove from NVM after identifying all unpopular keys in the chosen compaction key range.
The key popularity bitmap, \texttt{pop\_bitmap}, provides much faster access than looking up objects in the B-tree and the mapper. 
\texttt{Get}s set the key’s bitmap value to 1, while evictions from the tracker set the value to 0. 
Keys with value 0 are treated as cold objects and have a coldness value of 1.
This way, \texttt{pop\_map} approximates a key's  popularity and coldness score without retrieving the accurate clock value from the mapper. 
The \texttt{nvm\_bitmap} tracks keys present on NVM. \texttt{Put}s set the key's bit to 1, while compactions set it to 0.
The \texttt{flash\_bitmap} tracks if a key has any version (latest or older) present on flash. 
Compactions set the key's bit to 1, while \texttt{Delete}s set it to 0. 
The \texttt{nvm\_bitmap} and \texttt{flash\_bitmap} piece together information about overlapped keys. 
AND-ing the two bitmaps and counting the number of bit 1 gives the number of keys that exist on both NVM and flash.

Given a compaction key range, we sum over weighted parameters from each overlapped bucket to estimate the values of $p$, $o$, $F$, and coldness, and to compute an MSC score for the compaction key range.
The weight of each bucket equals to the ratio of the overlapped region to bucket size.
In Figure~\ref{fig:impl-msc}, Bucket 1 has 75\% of its key space overlapping with the compaction key range ([25, 99] shown in grey). 
Bucket 2 has a 25\% overlap ([100,125] shown in yellow). 
Thus, the weight is 0.75 for Bucket 1, and 0.25 for Bucket 2. Suppose the \texttt{num\_uniq\_keys} in Bucket 1 is 400 and in Bucket 2 is 100, then the estimated number of unique keys on NVM for key range [25, 99] will be 300 (\(0.75 \times 400\)) and for key range [100, 125] will be 25 (\(0.25 \times 100\)).
The total number of NVM keys for the compaction key range is the sum of these weighted numbers from each overlapped bucket, which equals to 325 keys. Other compaction key range parameters are computed in the same manner.

\paragrapha{Isolation and consistency guarantees.}
\name guarantees atomicity of individual writes and provides the read-committed isolation level (default setting in RocksDB, PostgreSQL). The current version does not support atomic batched writes, transactions and snapshots. We leave these features for future work.

\paragrapha{Crash consistency and recovery.}
\name does not use a write-ahead log for crash recovery. Instead, client writes are committed synchronously to their slab locations on NVM.
Current implementation of \name only supports objects upto 4KB in size. 
Objects are always stored within a single page.
Since Optane drives can write 4KB pages atomically even in case of power failures, \name leverages this capability to perform atomic updates for these sub-page sized objects.
For drives that don't support atomic writes to pages, \name can be modified to write updated data in new slab locations. Old slot entries can be reclaimed as soon as the write to the new location succeeds. \name can employ this same technique to support objects larger than 4KB.


\name ensures crash consistency on NVM using a logical timestamp entry. 
It is part of the object metadata and is synchronously written to disk along with the object.
Each partition maintains its own logical timestamp, which is a counter that is incremented by every \texttt{Put} or \texttt{Delete} operation for that partition.
During recovery, \name scans over all the NVM slabs, and skips key entries with older timestamps when reconstructing the B-tree index in memory. 
Similar to RocksDB~\cite{rocksdb}, \name keeps an on-disk manifest file that stores the active list of partition SST files for generating a consistent view of the flash database. For keys that exist on both NVM and flash, \name treats the NVM version as the latest version. Finally, partitions run recovery in parallel since they contain disjoint key spaces and don't require any coordination. This further speeds up recovery.

\section{Evaluation}
\label{sec:eval}


\begin{table}[t!]
    \centering
    \footnotesize
    \begin{tabular}{ll}
      \toprule 
      Type & Description (read\%, update\%)\\
      \midrule
      write heavy & A (50\%, 50\%), F (50\%, 50\% read-modify-writes) \\
      read heavy & B (95\%, 5\%), C (100\%, 0\%), D (latest; 95\%, 5\%) \\
      scan heavy & E (95\% scans, 5\% updates) \\
      \bottomrule
    \end{tabular}
    \caption{YCSB workload description.}
    \vspace{-2ex}
    \label{tab:ycsb}
\end{table}

\begin{figure}[t!]
	\begin{center}
	{\includegraphics[width=0.7\columnwidth]{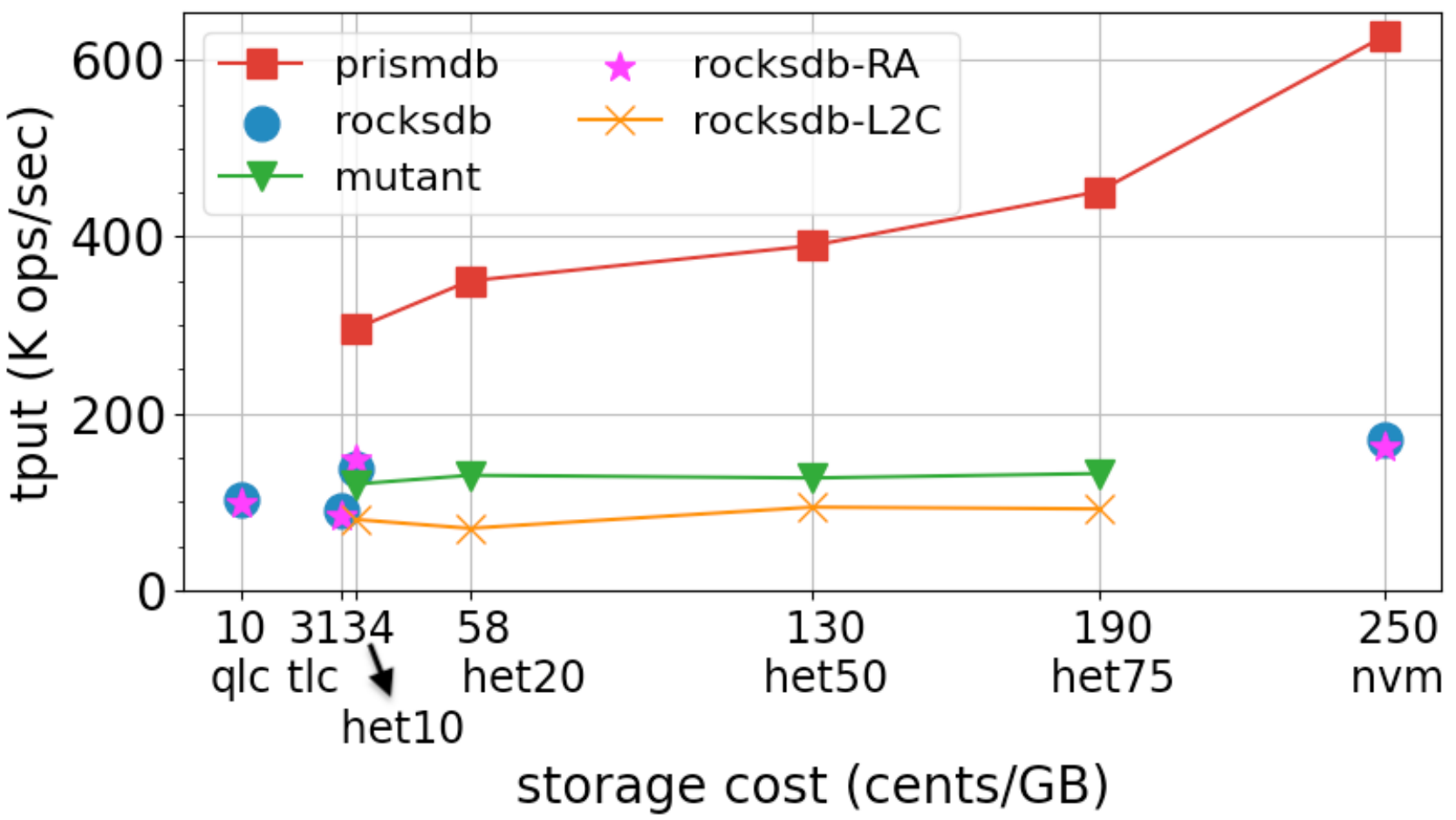}}
	\end{center}
	\vspace{-1ex}
	\caption{Throughput vs. storage cost under YCSB-A. ``hetX'' means heterogeneous or multi-tier setup with X\% on NVM.}
    \vspace{-3ex}
	\label{fig:main-tput}
\end{figure}


\begin{figure*}[!t]
	\centering
	\subfloat[]{
	\label{fig:het-tput}
		\includegraphics[width=0.26\textwidth]{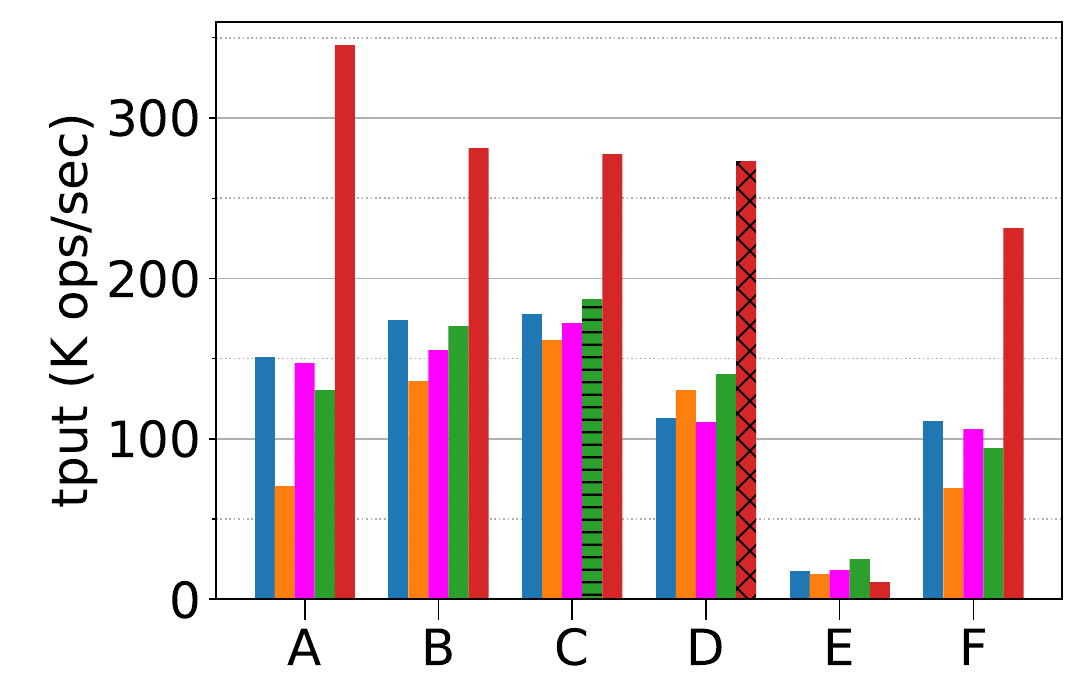}
	}~
	\subfloat[]{
	\label{fig:p50-latency}
		\includegraphics[width=0.31\textwidth]{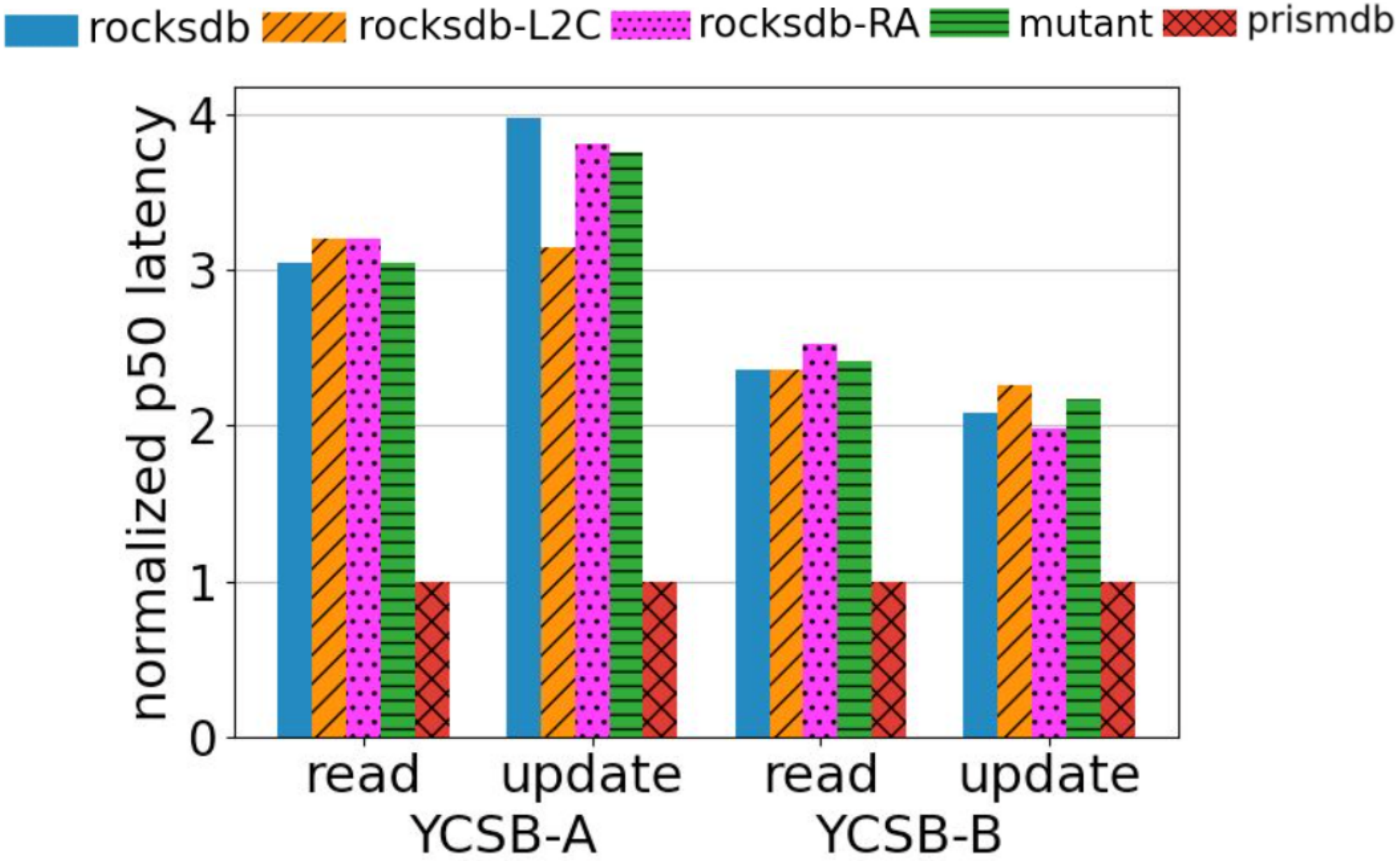}
	}~
	\subfloat[]{
	\label{fig:p99-latency}
		\includegraphics[width=0.23\textwidth]{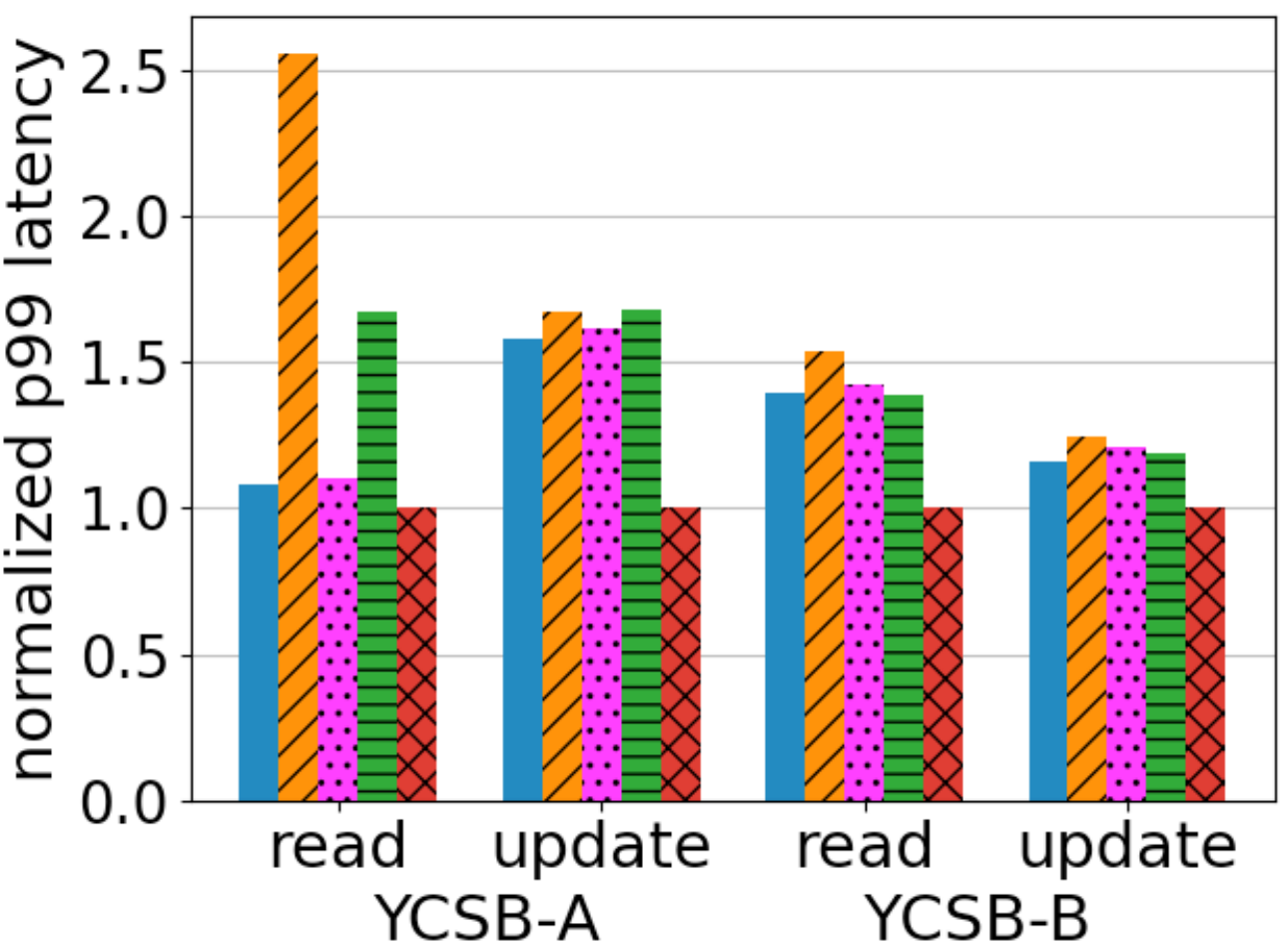}
	}
		\vspace{-1ex}
	\caption{(a) Throughput, (b) normalized median latency, and (c) normalized p99 latency.}
	\label{fig:ycsb-multi-tiered}
\end{figure*}


\begin{figure*}[!t]
	\centering
	\subfloat[]{
	\label{fig:dist-sweep-tput}
		\includegraphics[width=0.23\textwidth]{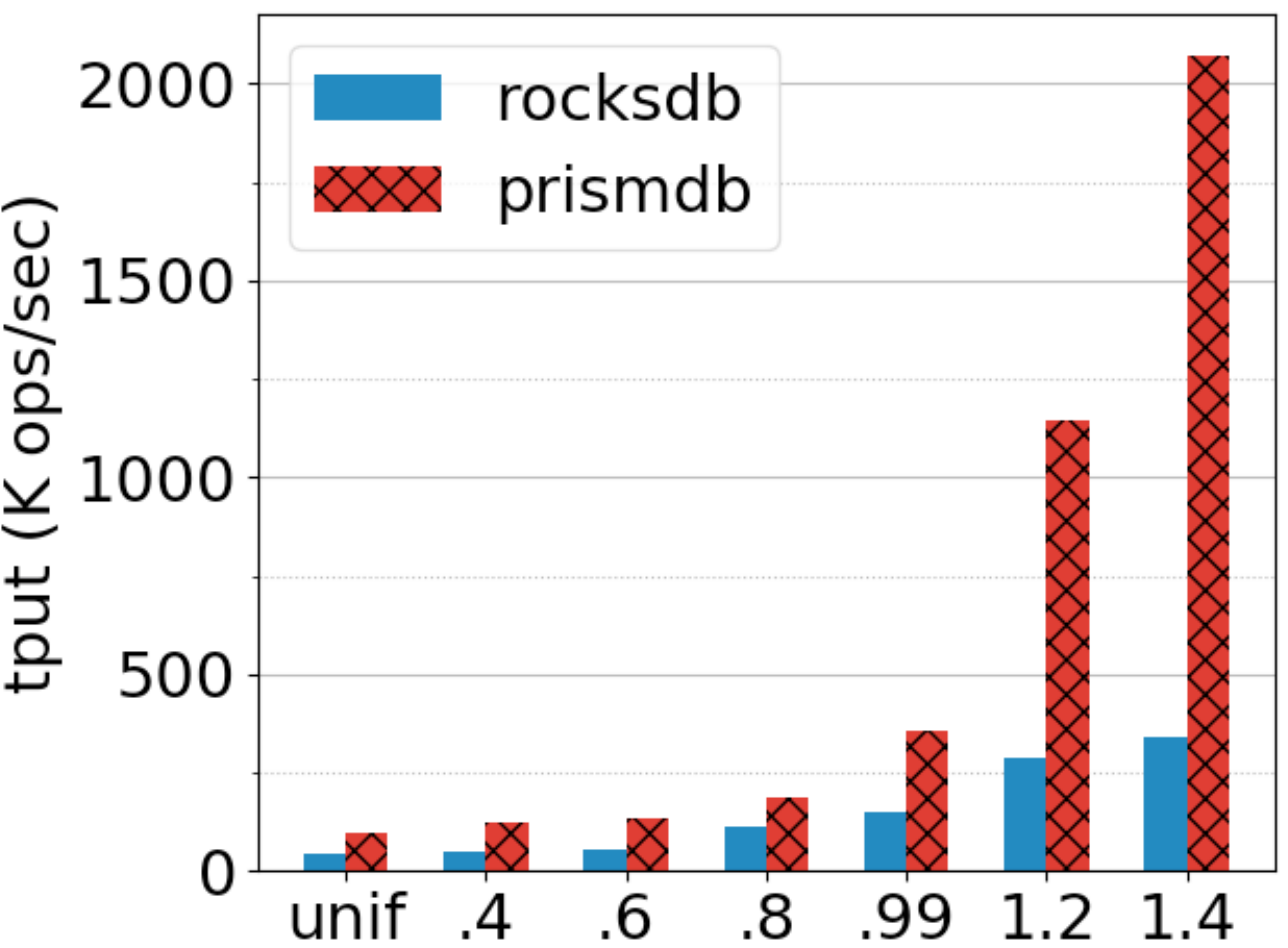}
	}~
	\subfloat[]{
	\label{fig:dist-sweep-rd-lat}
		\includegraphics[width=0.23\textwidth]{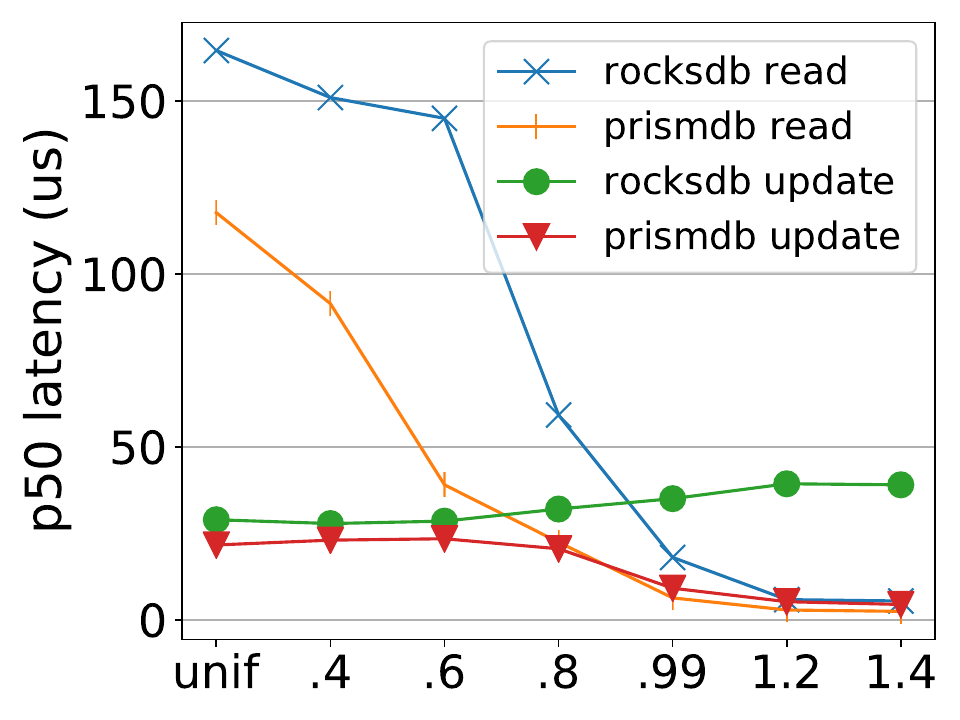}
	}~
	\subfloat[]{
	\label{fig:dist-sweep-wr-lat}
		\includegraphics[width=0.23\textwidth]{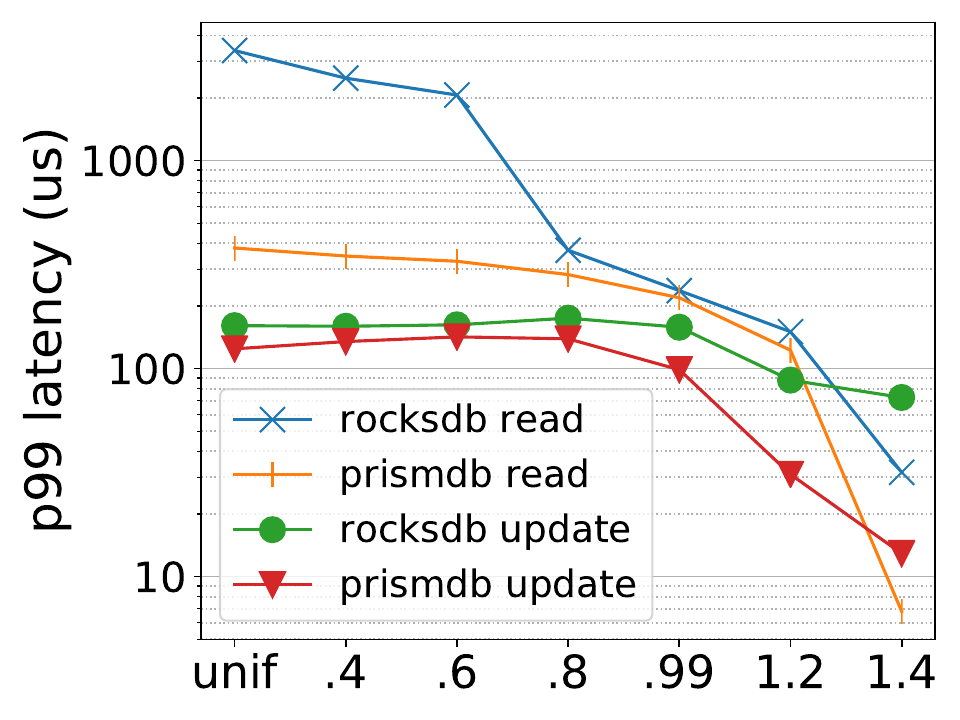}
	}
		\vspace{-1ex}
	\caption{YCSB-A performance with different Zipfian parameters.} 
	\label{fig:zipfians}
    \vspace{-2ex}
	\end{figure*}

We evaluate \name by answering the following questions:
 \begin{denseenum}
    \item How does \name compare to baselines (\S\ref{sec:homo-v-hetero})?
    \item Which workloads does \name benefit (\S\ref{sec:hetro-eval}, \S\ref{sec:prod})?
    \item How do compactions impact performance (\S\ref{sec:components})?
    \item How to set the pinning threshold (\S\ref{sec:components})?
 \end{denseenum}

\paragrapha{Configuration.}
We performed our experiments on a 32-core, 64~GB RAM machine running Ubuntu 18.04.2. Three different storage devices are locally attached to this machine via PCIe3.1: Intel Optane SSD P5800X (NVM), Intel 760p (TLC NAND), and Intel 660p (QLC NAND).  
All workloads are run using 8 concurrent clients restricted with cgroup to 10 CPU cores. By default all workloads use a 1:10 ratio between DRAM and storage, where 20\% of the DRAM is dedicated to block cache for LSM based systems (a common production configuration~\cite{rocksdbcidr}). The multi-tier configurations by default use a 1:5 size ratio between NVM and QLC.
For \name, we set the tracker size to 20\% of the total key space and the pinning threshold to 70\%. 
Unless otherwise specified, other settings are the default ones used in RocksDB.

\paragrapha{Workloads.}
We run YCSB (Table~\ref{tab:ycsb}) and three Twitter production workloads \cite{twitter-cache}. For YCSB, by default we use YCSB-A with Zipfian 0.99 distribution, 100~M key dataset and a fixed 1~KB object size, with a total database size of 100~GB. For the update-intensive workloads (A and F) the first half of the trace is used as a warm-up period. For read-intensive workloads (B, C and D), we run 300~M requests with a longer warm-up (80\%) to allow the promotion-based compactions to take effect. The scan workload is run with 10~M requests and a 50\% warm-up.

We choose three representative Twitter traces~\cite{twitter-cache} with varying reads to write ratios and distributions: write heavy (cluster39), mixed (cluster19) and read heavy (cluster51). Write heavy trace has 6:94 read  write ratio and uniform writes. Mixed trace has 75:25 read write ratio, with zipfian reads and uniform writes. Read intensive trace has 90:10 read write ratio and a zipfian access pattern.



\paragrapha{Baselines.} 
We compare \name against five baselines: RocksDB (v6.2.0), RocksDB that uses NVM as an L2 read cache (labeled rocksdb-l2c), our initial read-aware LSM prototype from \S\ref{sec:perf} (labeled as rocksdb-RA), and two academic LSM KV stores, Mutant~\cite{mutant} and SpanDB~\cite{spandb}. 
Mutant is a storage layer for LSMs that tracks access popularity of SST files and places them across heterogeneous storage accordingly. SpanDB uses SPDK~\cite{spdk} on the Optane drive to bypass the kernel's I/O overheads. Aside of its use of SPDK, SpanDB's data layout and compaction are nearly identical to RocksDB's. For a fair comparison, we restrict SpanDB's auto placement module to use the NVM:QLC ratio size of 1:5. Since SpanDB by design bypasses the page cache and persists all WAL writes to disk synchronously, we evaluate it separately by enabling fsync mode in RocksDB and \name (RocksDB by default does not fsync the WAL).

We also considered comparing against PebblesDB~\cite{pebblesdb}, an LSM-based KV store that trades read performance for write throughput. However, in our experiments, RocksDB consistently outperformed a tuned PebblesDB on all YCSB workloads 
(this matches the findings of others~\cite{evendb}). On top of that, PebblesDB does not have support for tiered storage, so we do not use it as a baseline. 




\subsection{Single-tier vs. Multi-tier}
\label{sec:homo-v-hetero}

Figure~\ref{fig:main-tput} compares the average throughput and storage cost between \name and baseline systems under seven configurations: three single-tier configurations (NVM, TLC, and QLC) and four multi-tier configuration (het). Note that since default RocksDB places data in different storage types on a level granularity, we cannot create a configuration that will match every point on the X axis with a fixed LSM tree shape.


As expected, for regular RocksDB, NVM outperforms single-tier TLC NAND and QLC NAND setup, which uses denser and slower flash. Surprisingly, the QLC setup slightly outperforms TLC, which we attribute to the internal cache on the newer QLC device. Across the multi-tier setups, the higher the proportion of NVM, the better the performance.
Interestingly, the L2 cache configuration of RocksDB consistently underperforms the level-by-level configuration. The reason for this is that the L2 cache serves as a read cache, but not a write cache. All writes go to flash, which is slow. Additionally, the L2 cache needs to load frequently updated objects from flash to NVM which degrades system throughput.


\name significantly outperforms the baselines on all multi-tier storage setups. Notably, the \emph{het10} configuration with \name has 3.3$\times$ better throughput and 2$\times$ lower tail latency than RocksDB with pure TLC, which is the default way RocksDB is deployed in datacenters today, while costing almost the same (\$0.34/GB vs. \$0.31/GB). Note these prices fluctuate on a daily basis.
Mutant's performance is equivalent to RocksDB's, because it maps popular files to NVM, this comes at the cost of triggering more compactions, especially for write-dominated workloads like YCSB-A. For example, Mutant incurs 59\% more background compactions on this workload.
In addition, Mutant's mapping is coarse grained, since it makes placement decisions on a file-by-file basis, and a single file may have objects with varying popularity. 
Read aware RocksDB suffers from the trade-off between pinning and compaction efficiency. Its read performance gains from serving more reads from NVM are negated by increased compactions, thus offering little to no throughput or latency advantage.

\subsection{Multi-tier Storage}
\label{sec:hetro-eval}

\paragrapha{YCSB Sweep.} Figure~\ref{fig:ycsb-multi-tiered} compares the throughput, median and 99th percentile latencies across the different YCSB workloads. Mutant slightly outperforms RocksDB on read-friendly workloads (YCSB-C and YCSB-D) because it can map popular SST files to NVM better under low compaction churn. RocksDB's L2 cache configuration performs better than RocksDB in limited workloads (YCSB-D only). \name consistently outperforms all baselines for point queries, in terms of throughput and tail latency, due to its more efficient data layout and compaction algorithm. Notably, it is able to outperform the baselines even for YCSB-B and YCSB-C, which are read-heavy and read-only, respectively. In these workloads, \name's promotions get triggered to migrate hot data that may have been compacted in the past to flash. However, \name offers the biggest performance improvement for workloads that include a significant percentage of writes (\eg YCSB-A). 
For that workload, \name saves more than 1.9$\times$ CPU and does more useful work compared to LSM-based key-value stores that still need to run traditional compactions on NVM.

The only YCSB workload where \name does not outperform the baselines is the scan workload. RocksDB in particular is optimized for scans, and includes a prefetcher that proactively fetches blocks, which greatly improves its performance for predictable scan patterns. When we disable RocksDB's prefetcher, we find both systems exhibit comparable performance under scans. We leave implementing a prefetcher for \name as future work. Moreover, the scan workload does not have enough access popularity to exploit. YCSB-E picks the start key of a scan query from a zipfian distribution and selects a random scan length. This over-simplifies scan patterns seen in real world workloads~\cite{fb2020} that exhibit strong key-space locality where hot keys are closely located in the key-space, forming dense key-range hot spots. \name's core techniques of tracking key hotness and maintaining hot-cold separation across storage tiers should benefit real-world workloads with good key-range locality.


\paragrapha{Data skewness.} 
Next, we evaluate how data skewness impacts \name.   Figure~\ref{fig:zipfians} presents the results of a key distribution sweep using YCSB-A. \name is able to provide a throughput benefit under all distributions. 
For highly skewed workloads, due to popularity tracking and data pinning, \name serves more reads from DRAM or NVM and absorbs more writes as in-place updates in NVM, delivering superior performance.
For uniform workloads, due to \name's partitioned design and NVM data layout, it has much lower tail latency compared to RocksDB. RocksDB does excessive compactions across levels which are known to severely impact tail latency\cite{SILK}.



\begin{figure*}[!t]
	\centering
	\subfloat[]{
	\label{fig:read-lat-cdf}
		\includegraphics[width=0.23\textwidth]{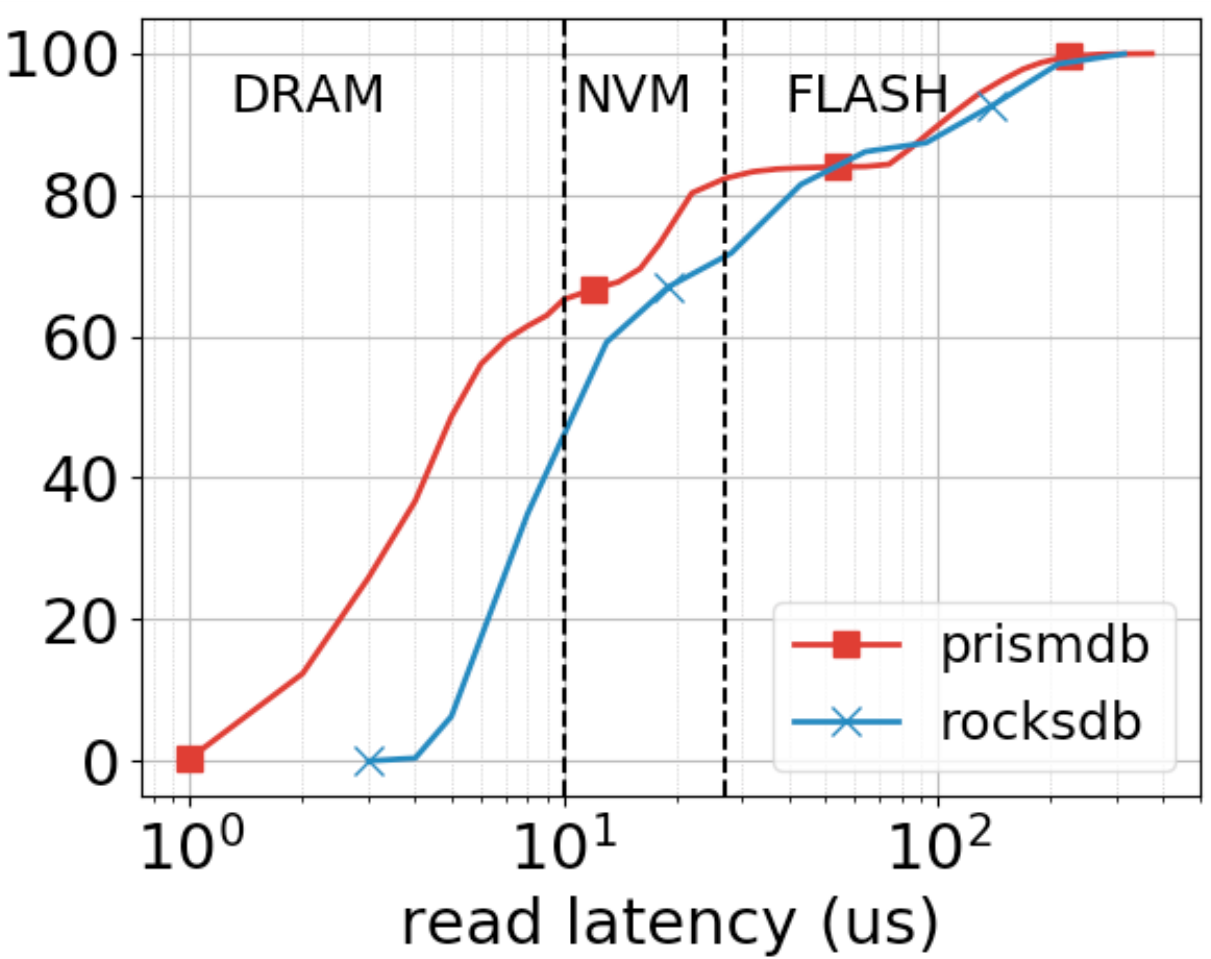}
	}
	\subfloat[]{
	\label{fig:promotions-tput}
		\includegraphics[width=0.25\textwidth]{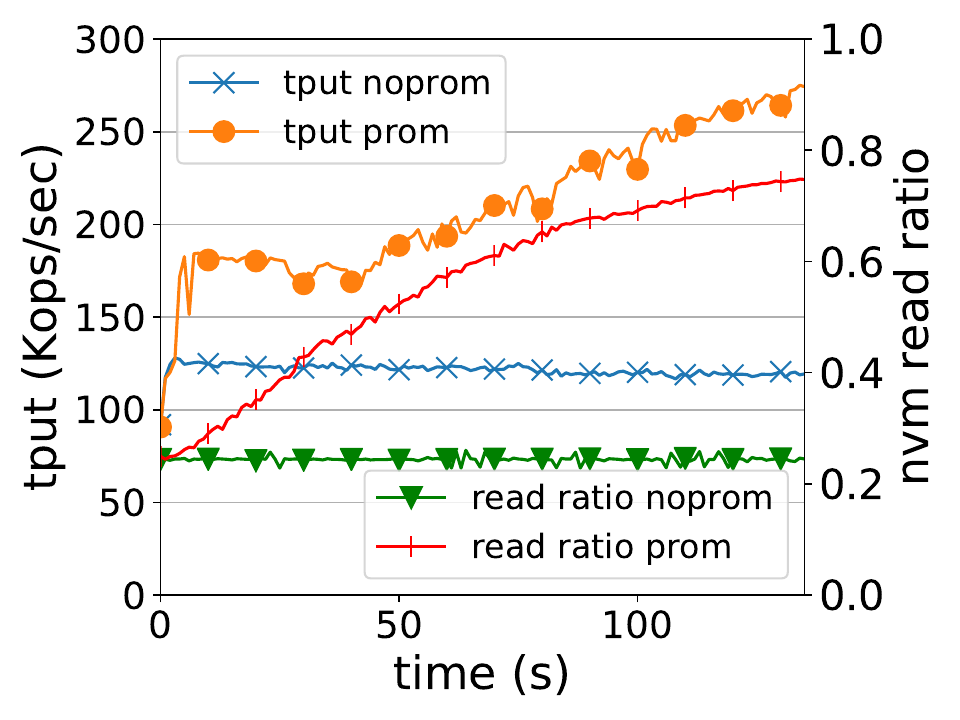}
	}
	\subfloat[]{
	\label{fig:pinning-threshold}
		\includegraphics[width=0.22\textwidth]{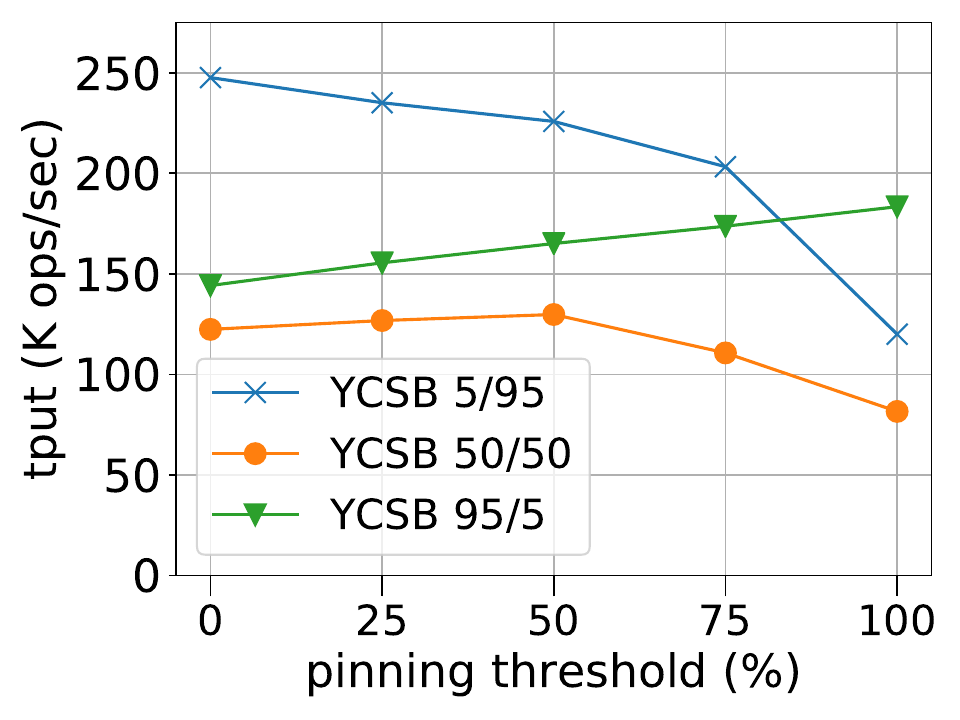}
	}
	\subfloat[]{
	\label{fig:partition-scaling}
		\includegraphics[width=0.22\textwidth]{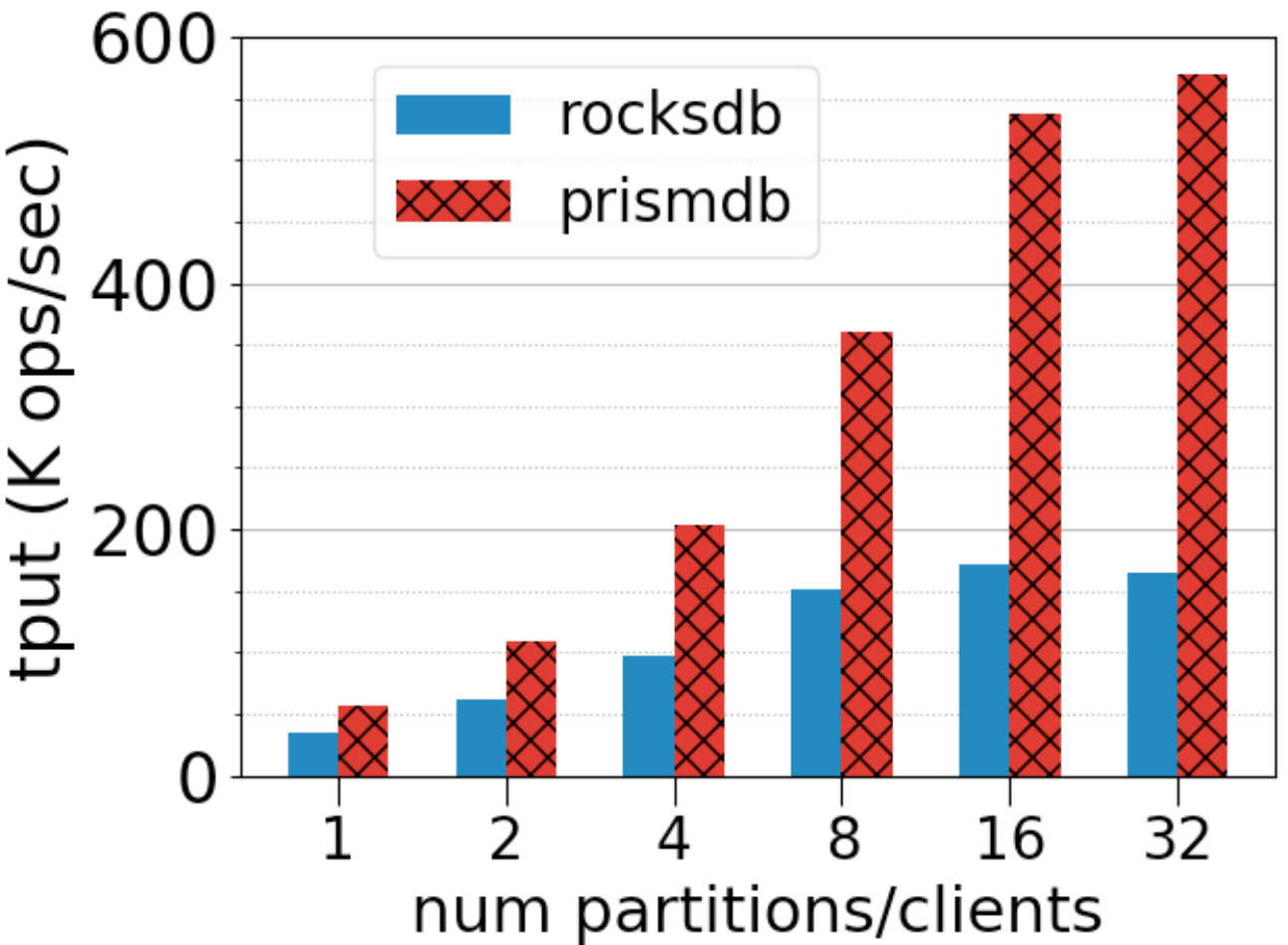}
	}
	\vspace{-1ex}
	\caption{(a) Read latency cdf on YCSB-B. (b)  Impact of promotions on read-only YCSB-C. (c) Impact of pinning threshold. (d) Throughput as function of number partitions/clients.} 
	\vspace{-3ex}
\end{figure*}

To better understand why \name provides better read latency, we plot a CDF comparing the latencies of \name to RocksDB on read-heavy YCSB-B (Figure~\ref{fig:read-lat-cdf}). As expected, the figure shows that \name serves a lower percentage of its requests from flash compared to RocksDB due to better data placement. Interestingly, \name also more efficiently utilizes DRAM. Over time, NVM blocks in \name become more densely packed with popular objects, thus improving the hit rate of the OS page cache. In contrast, RocksDB caches data blocks in DRAM with mixed popularity and further ``pollutes'' much of its DRAM cache with compactions.

\begin{figure}[!t]
	\centering
	\subfloat[]{
	\label{fig:fsync-tput}
		\includegraphics[width=0.42\columnwidth]{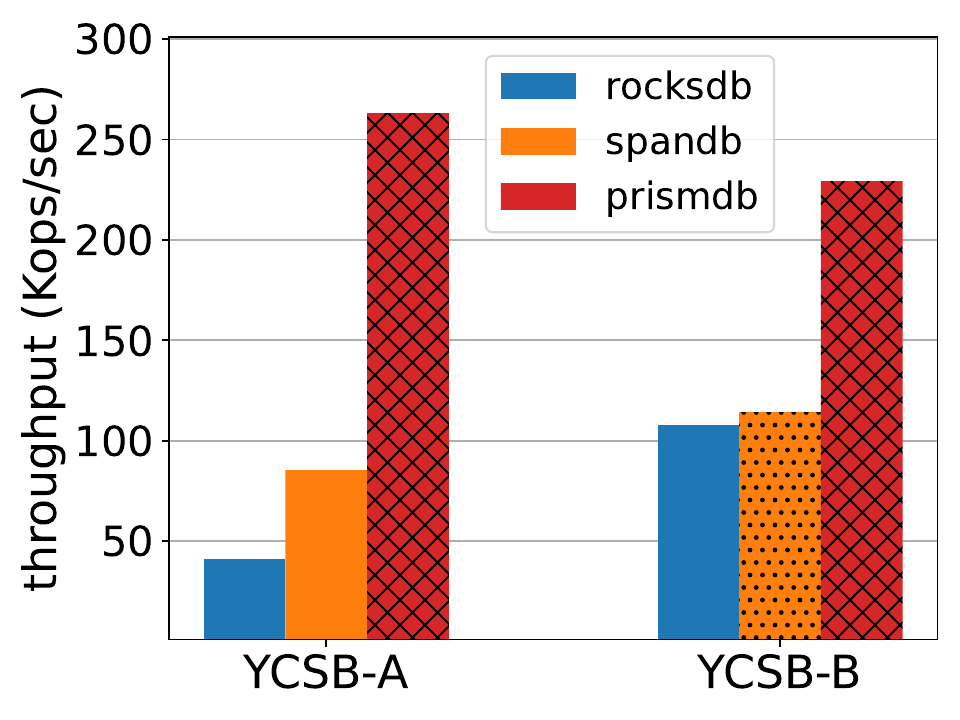}
	}~
	\subfloat[]{
	\label{fig:fsync-lat}
		\includegraphics[width=0.43\columnwidth]{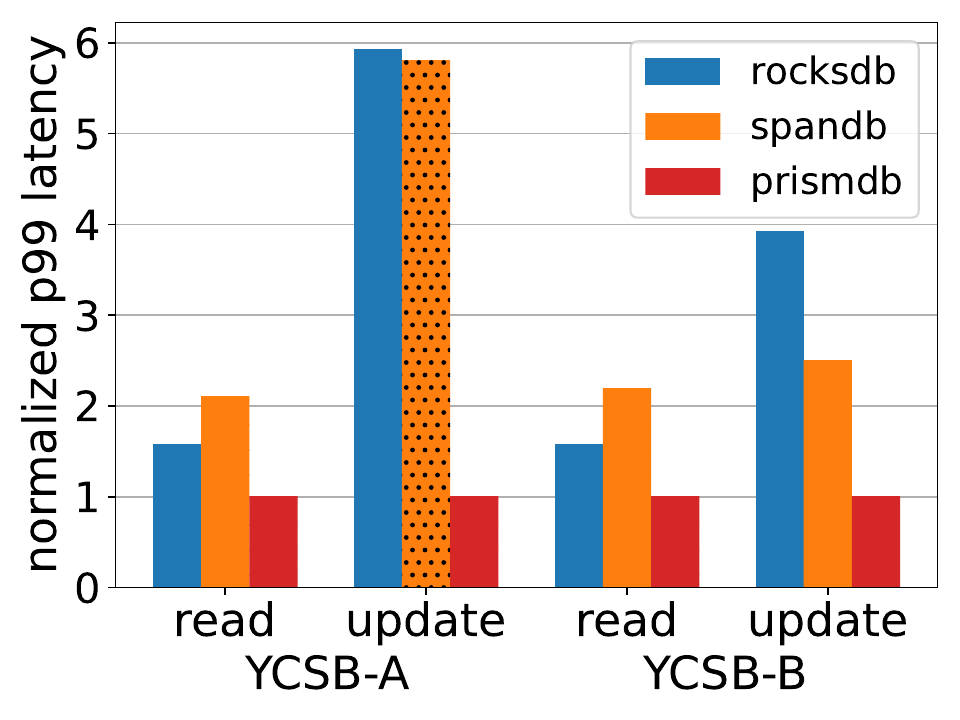}
	}
    \vspace{-1ex}
	\caption{Performance with fsync enabled.}
	\vspace{-3ex}
	\label{fig:het}
\end{figure}

\paragrapha{Performance with fsync.}
SpanDB is specifically designed to provide fast synchronous WAL logging for write heavy workloads. Figure~\ref{fig:fsync-tput} compares \name with RocksDB and SpanDB with fsync WAL mode enabled. \name outperforms RocksDB by 6.4~$\times$ and SpanDB by 3~$\times$ respectively. RocksDB's group commit protocol creates a bottleneck for logging on Optane, which SpanDB alleviates by using SPDK for logging. However, \name's partitioned approach completely removes the centralized logging bottleneck. Further, like RocksDB, SpanDB spends significant CPU resources on compactions to keep data sorted on LSM levels resident on Optane. Due to its use of SPDK, SpanDB also requires the application to dedicate cores to busy-poll on I/O, which is inefficient when CPU is a bottleneck. \name's data layout on Optane and efficient migrations achieve much lower median and tail latency compared to RocksDB and SpanDB.

\paragrapha{QLC Endurance.}
So far, we have presented storage costs in terms of \$/GB of capacity. But for lower endurance dense flash (\eg QLC NAND), the application write-rate is another factor that determines the total cost of ownership (TCO), in case the drive wears out and needs to be replaced before its intended lifetime (typically 3-5 years).
Thus, the lifetime of \name heavily depends on how quickly its flash component wears out.

\begin{figure}[t!]
	\begin{center}
	{\includegraphics[width=0.6\columnwidth]{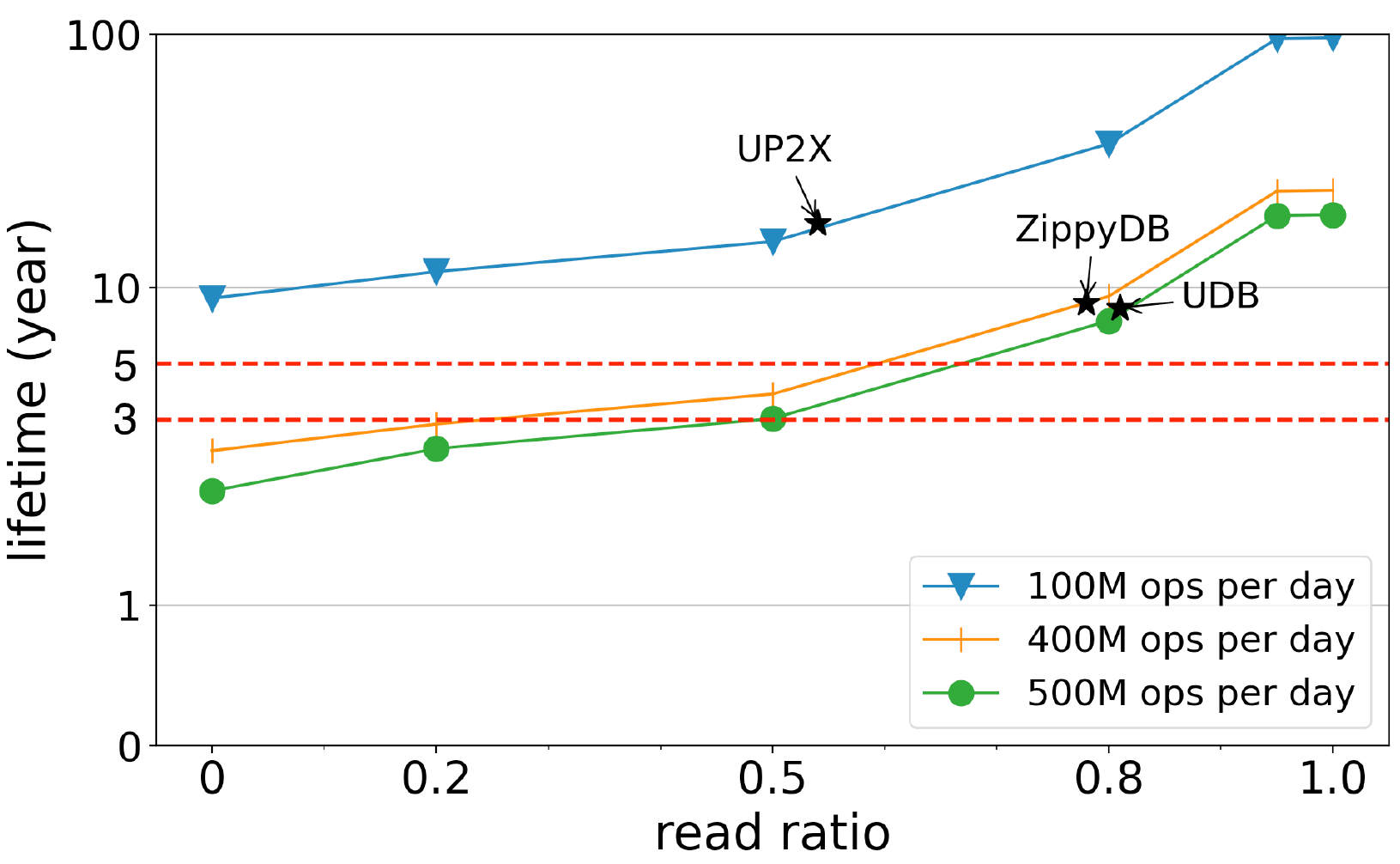}}
	\end{center}
    \vspace{-1ex}
	\caption{\name's lifetime under different workloads}
    \vspace{-3ex}
	\label{fig:endurance}
\end{figure}

Figure~\ref{fig:endurance} evaluates the lifetime of \name under different workload settings. 
We assume a reasonable DB size seen in real deployment~\cite{rocksdbcidr}, 600GB, and use the load numbers reported from production systems~\cite{fb2020} to compute the lifetime of \name over a range of read/write ratios. In addition, we select three large-scale applications (\ie UP2X, ZippyDB, and UDB) as reported~\cite{fb2020} and annotate their projected lifetime under \name in Figure~\ref{fig:endurance}. Most real-world workloads are read-dominated (\eg 99.8\% requests in TAO are reads)~\cite{spanner, tao, fb2020}.
\name with QLC flash is a good fit for such systems.

Update-heavy workloads with high request rate, on the other hand, will wear out the QLC flash before the 3-5 year period, and thus incur a higher TCO. For such workloads, system administrators can consider using higher endurance flash like TLC NAND.


\subsection{Twitter Production Workloads}
\label{sec:prod}

\begin{table}
    \centering
    \footnotesize 
    \begin{tabular}{l | r r | r r }
      \multicolumn{1}{c}{} & \multicolumn{2}{c}{Throughput (ops/s)} & \multicolumn{2}{c}{Avg put latency}\\
      \multicolumn{1}{c|}{Trace} & RocksDB & \name & RocksDB & \name \\
      \midrule
      write-heavy (cluster39) & 236K  & 337K & 35.3 $\mu$s & 24.6 $\mu$s\\
      mixed (cluster19) & 245K & 248K & 25.4 $\mu$s & 15.0 $\mu$s\\
      read-heavy (cluster51) & 798K & 2,620K & 22.9 $\mu$s & 5.4 $\mu$s\\
      \bottomrule
    \end{tabular}
    \caption{Performance on Twitter workloads.}
    \vspace{-3ex}
    \label{tab:twitter}
\end{table}

Table~\ref{tab:twitter} compares \name with RocksDB on three Twitter production workloads, which correspond to three different workload patterns. \name has 43\% higher throughput for the insert-heavy workload (cluster39), due to its optimized writes and compactions. On the mixed workload (cluster19) it does not show any improvements. We identify two reasons for this. First, cluster19 workload has highly cacheable reads that fit in memory, so \name's read optimizations don't have any impact. Second, this trace contains tiny objects (102~B avg size) which creates an interesting performance challenge, because of \name's reliance on the OS page cache, which reads from or writes to NVM at a 4~KB granularity.
We therefore added a small optimization that sorts free NVM slab slots by their disk locations, which ensures consecutive writes go to the same OS page.
However compacting tiny objects still incurs 500~K random NVM reads which make compactions slower, resulting in the overall throughput of \name to be similar to RocksDB's.
\name can be further optimized for tiny objects (sizes less than 128B) by using direct IO on Optane and bypassing the OS page cache. We leave the optimization for tiny objects as future work. On read heavy workload (cluster51), \name exhibits a 3.2$\times$ improvement in throughput due to larger objects (370~B avg size) and Zipfian reads and writes, for which \name's read and write optimizations are important.

\subsection{Evaluating System Components}
\label{sec:components}

We now evaluate different components of \name.
Figure~\ref{fig:promotions-tput} shows the effect of promotions (moving data from QLC to NVM) under a read-only workload (YCSB-C). The figure shows that promotions significantly increase the ratio of reads from NVM, resulting in higher overall throughput. 


Figure~\ref{fig:pinning-threshold} plots the throughput of three YCSB workloads as a percentage of the pinning threshold. In this experiment, the tracker records popularity for 20\% of the total objects, and the pinning threshold is calculated as a percentage of the tracker size. The experiment shows that the ideal pinning threshold is different for each workload. For a read-dominated workload, a higher threshold is better, since it is more important to retain popular keys even at the expense of less-efficient garbage collection. As expected, the reverse is true for write-dominated workloads. Automatically choosing the right pinning threshold for dynamic workloads can be done via a hill climbing approach, which we leave for future work. 

Figure~\ref{fig:partition-scaling} shows how \name scales as a function of partitions (or clients). The figure shows that \name continues to scale as a function of the number of partitions, until it exceeds 10 partitions, which is equal to the number of cores in our experiment.

\subsection{Evaluation Summary}
\label{sec:summary}

Our evaluation demonstrates the benefits of using multiple tiers of storage, and demonstrates that \name's design, which is tailored to multiple storage tiers, is superior to existing LSM-based systems. 
\name outperforms other baselines across most workloads, both synthetic and real-world, under both relaxed and strict durability requirements (\ie with fsync enabled). 
In particular, \name's techniques for popularity tracking and object pinning improve both read and write performance in Zipfian workloads. In addition, \name's novel cost-benefit compaction metric (MSC) achieves better hot-cold separation across storage tiers without incurring high write amplification on flash. We show that even a low-endurance flash drive like QLC NAND, when used in a multi-tier configuration under \name, can meet the 3-5 years lifetime requirement for many common datacenter workloads.

\paragrapha{Limitations.} \name has a few limitations:
(1) Write heavy workloads with high request loads are not suitable for the lower endurance QLC NAND. For such workloads, \name should be provisioned with higher endurance flash;
(2) Current version of \name has sub-optimal scan performance than other baselines due to lack of a sophisticated prefetcher;
(3) Tiny objects (sizes less than 128B) incur high read amplification during compactions since compacting a single NVM object requires reading a 4KB page. Compacting a large number of tiny objects pollutes the OS page cache and degrades overall system performance. One possible solution is to implement an application level page cache and bypass the OS page cache (not yet implemented in \name).

\section{Other Related Work}

We reviewed closely related work in \S\ref{sec:bg}. We now describe other related work. 
SplinterDB~\cite{splinterdb}, KVell~\cite{KVell},
and uDepot~\cite{udepot} are databases designed for fast NVMe devices (\eg Optane SSD). Since they are optimized for NVM, they can avoid the buffering and large compactions employed by LSMs. However, they would not perform well in a tiered setting that includes traditional flash, which requires large contiguous writes. 
In addition, some key-value stores use persistent memory as a second byte-addressable memory tier~\cite{write-behind,arulrajsigmod,ogleari2018steal,vanrenen2,HiKV,matrixkv,vanrenen,kaiyrakhmet2019slm,persistentmemcached,hilsm,akram2021exploiting,spitfire}. We focus on tiered block-addressable storage compaction, and our techniques are complementary to a multi-tiered memory setting. 
Strata~\cite{kwon2017strata} is a file system for multi-tiered storage that places data at a file granularity. 
As our comparison with Mutant demonstrates, managing data placement at the file level is not suitable for datacenter key-value stores, in which each file may contain thousands of small objects of varying popularity.

\section{Conclusion}

By combining multiple storage technologies within the same key-value store we can enable both \emph{fast} and \emph{affordable} storage engines. To achieve these goals, we demonstrated the importance of designing new hybrid data structures and compaction mechanisms, which are tailored for multi-tier storage setup. 
We believe that the general approach of simultaneously employing a mixture of storage technologies will likely prove useful for other areas of systems.



\label{lastpage}

\bibliographystyle{ACM-Reference-Format}
\bibliography{main}


\begin{thebibliography}{00}


\ifx \showCODEN    \undefined \def \showCODEN     #1{\unskip}     \fi
\ifx \showDOI      \undefined \def \showDOI       #1{#1}\fi
\ifx \showISBNx    \undefined \def \showISBNx     #1{\unskip}     \fi
\ifx \showISBNxiii \undefined \def \showISBNxiii  #1{\unskip}     \fi
\ifx \showISSN     \undefined \def \showISSN      #1{\unskip}     \fi
\ifx \showLCCN     \undefined \def \showLCCN      #1{\unskip}     \fi
\ifx \shownote     \undefined \def \shownote      #1{#1}          \fi
\ifx \showarticletitle \undefined \def \showarticletitle #1{#1}   \fi
\ifx \showURL      \undefined \def \showURL       {\relax}        \fi
\providecommand\bibfield[2]{#2}
\providecommand\bibinfo[2]{#2}
\providecommand\natexlab[1]{#1}
\providecommand\showeprint[2][]{arXiv:#2}

\bibitem[\protect\citeauthoryear{??}{fio}{[n. d.]}]%
        {fio}
 \bibinfo{year}{[n. d.]}\natexlab{}.
\newblock \bibinfo{title}{Flexible {I/O} Tester}.
\newblock \bibinfo{howpublished}{https://github.com/axboe/fio}.
  (\bibinfo{year}{[n. d.]}).
\newblock


\bibitem[\protect\citeauthoryear{??}{opt}{[n. d.]}]%
        {optane}
 \bibinfo{year}{[n. d.]}\natexlab{}.
\newblock \bibinfo{title}{Intel {O}ptane memory}.
\newblock   (\bibinfo{year}{[n. d.]}).
\newblock
\newblock
\shownote{\url{https://www.intel.com/content/www/us/en/architecture-and-technology/optane-memory.html}.}


\bibitem[\protect\citeauthoryear{??}{lev}{[n. d.]}]%
        {leveldb}
 \bibinfo{year}{[n. d.]}\natexlab{}.
\newblock \bibinfo{title}{Level{DB}}.
\newblock   (\bibinfo{year}{[n. d.]}).
\newblock
\newblock
\shownote{\url{http://leveldb.org/}.}


\bibitem[\protect\citeauthoryear{??}{zna}{[n. d.]}]%
        {znand}
 \bibinfo{year}{[n. d.]}\natexlab{}.
\newblock \bibinfo{title}{Samsung {Z-SSD} Redefining fast responsiveness}.
\newblock   (\bibinfo{year}{[n. d.]}).
\newblock
\newblock
\shownote{\url{https://www.samsung.com/semiconductor/ssd/z-ssd/}.}


\bibitem[\protect\citeauthoryear{Akram}{Akram}{2021}]%
        {akram2021exploiting}
\bibfield{author}{\bibinfo{person}{Shoaib Akram}.}
  \bibinfo{year}{2021}\natexlab{}.
\newblock \showarticletitle{Exploiting Intel optane persistent memory for full
  text search}. In \bibinfo{booktitle}{{\em Proceedings of the 2021 ACM SIGPLAN
  International Symposium on Memory Management}}. \bibinfo{pages}{80--93}.
\newblock


\bibitem[\protect\citeauthoryear{Arulraj and Pavlo}{Arulraj and Pavlo}{2017}]%
        {arulrajsigmod}
\bibfield{author}{\bibinfo{person}{Joy Arulraj} {and} \bibinfo{person}{Andrew
  Pavlo}.} \bibinfo{year}{2017}\natexlab{}.
\newblock \showarticletitle{How to Build a Non-Volatile Memory Database
  Management System}. In \bibinfo{booktitle}{{\em Proc. Intl. Conference on
  Management of Data (SIGMOD)}}. 6.
\newblock
\showISBNx{978-1-4503-4197-4}
\showDOI{%
\url{https://doi.org/10.1145/3035918.3054780}}


\bibitem[\protect\citeauthoryear{Arulraj, Perron, and Pavlo}{Arulraj
  et~al\mbox{.}}{2016}]%
        {write-behind}
\bibfield{author}{\bibinfo{person}{Joy Arulraj}, \bibinfo{person}{Matthew
  Perron}, {and} \bibinfo{person}{Andrew Pavlo}.}
  \bibinfo{year}{2016}\natexlab{}.
\newblock \showarticletitle{Write-behind Logging}.
\newblock \bibinfo{journal}{{\em Proc. VLDB Endowment\/}} \bibinfo{volume}{10},
  \bibinfo{number}{4} (\bibinfo{date}{Nov.} \bibinfo{year}{2016}), 12.
\newblock
\showISSN{2150-8097}
\showDOI{%
\url{https://doi.org/10.14778/3025111.3025116}}


\bibitem[\protect\citeauthoryear{Balmau, Didona, Guerraoui, Zwaenepoel, Yuan,
  Arora, Gupta, and Konka}{Balmau et~al\mbox{.}}{2017}]%
        {triad}
\bibfield{author}{\bibinfo{person}{Oana Balmau}, \bibinfo{person}{Diego
  Didona}, \bibinfo{person}{Rachid Guerraoui}, \bibinfo{person}{Willy
  Zwaenepoel}, \bibinfo{person}{Huapeng Yuan}, \bibinfo{person}{Aashray Arora},
  \bibinfo{person}{Karan Gupta}, {and} \bibinfo{person}{Pavan Konka}.}
  \bibinfo{year}{2017}\natexlab{}.
\newblock \showarticletitle{{TRIAD}: Creating Synergies Between Memory, Disk
  and Log in Log Structured Key-Value Stores}. In \bibinfo{booktitle}{{\em
  Proc. {USENIX} Annual Technical Conference (ATC)}}.
\newblock
\showISBNx{978-1-931971-38-6}
\showURL{%
\url{https://www.usenix.org/conference/atc17/technical-sessions/presentation/balmau}}


\bibitem[\protect\citeauthoryear{Balmau, Dinu, Zwaenepoel, Gupta,
  Chandhiramoorthi, and Didona}{Balmau et~al\mbox{.}}{2019}]%
        {SILK}
\bibfield{author}{\bibinfo{person}{Oana Balmau}, \bibinfo{person}{Florin Dinu},
  \bibinfo{person}{Willy Zwaenepoel}, \bibinfo{person}{Karan Gupta},
  \bibinfo{person}{Ravishankar Chandhiramoorthi}, {and} \bibinfo{person}{Diego
  Didona}.} \bibinfo{year}{2019}\natexlab{}.
\newblock \showarticletitle{{SILK}: Preventing Latency Spikes in Log-Structured
  Merge Key-Value Stores}. In \bibinfo{booktitle}{{\em Proc. {USENIX} Annual
  Technical Conference (ATC)}}.
\newblock
\showISBNx{978-1-939133-03-8}
\showURL{%
\url{https://www.usenix.org/conference/atc19/presentation/balmau}}


\bibitem[\protect\citeauthoryear{Beckmann, Chen, and Cidon}{Beckmann
  et~al\mbox{.}}{2018}]%
        {LHD}
\bibfield{author}{\bibinfo{person}{Nathan Beckmann}, \bibinfo{person}{Haoxian
  Chen}, {and} \bibinfo{person}{Asaf Cidon}.} \bibinfo{year}{2018}\natexlab{}.
\newblock \showarticletitle{{LHD}: Improving Cache Hit Rate by Maximizing Hit
  Density}. In \bibinfo{booktitle}{{\em Proc. {USENIX} Symposium on Networked
  Systems Design and Implementation ({NSDI})}}.
\newblock
\showISBNx{978-1-939133-01-4}
\showURL{%
\url{https://www.usenix.org/conference/nsdi18/presentation/beckmann}}


\bibitem[\protect\citeauthoryear{Bloom}{Bloom}{1970}]%
        {bloomfilter}
\bibfield{author}{\bibinfo{person}{Burton~H. Bloom}.}
  \bibinfo{year}{1970}\natexlab{}.
\newblock \showarticletitle{Space/time trade-offs in hash coding with allowable
  errors}.
\newblock \bibinfo{journal}{{\it Commun. ACM}} (\bibinfo{year}{1970}).
\newblock


\bibitem[\protect\citeauthoryear{Bronson, Amsden, Cabrera, Chakka, Dimov, Ding,
  Ferris, Giardullo, Kulkarni, Li, Marchukov, Petrov, Puzar, Song, and
  Venkataramani}{Bronson et~al\mbox{.}}{2013}]%
        {tao}
\bibfield{author}{\bibinfo{person}{Nathan Bronson}, \bibinfo{person}{Zach
  Amsden}, \bibinfo{person}{George Cabrera}, \bibinfo{person}{Prasad Chakka},
  \bibinfo{person}{Peter Dimov}, \bibinfo{person}{Hui Ding},
  \bibinfo{person}{Jack Ferris}, \bibinfo{person}{Anthony Giardullo},
  \bibinfo{person}{Sachin Kulkarni}, \bibinfo{person}{Harry Li},
  \bibinfo{person}{Mark Marchukov}, \bibinfo{person}{Dmitri Petrov},
  \bibinfo{person}{Lovro Puzar}, \bibinfo{person}{Yee~Jiun Song}, {and}
  \bibinfo{person}{Venkat Venkataramani}.} \bibinfo{year}{2013}\natexlab{}.
\newblock \showarticletitle{{TAO}: {F}acebook's {D}istributed {D}ata {S}tore
  for the {S}ocial {G}raph}. In \bibinfo{booktitle}{{\em Proc. USENIX Annual
  Technical Conference (ATC)}}.
\newblock
\showISBNx{978-1-931971-01-0}


\bibitem[\protect\citeauthoryear{Cao, Dong, Vemuri, and Du}{Cao
  et~al\mbox{.}}{2020}]%
        {fb2020}
\bibfield{author}{\bibinfo{person}{Zhichao Cao}, \bibinfo{person}{Siying Dong},
  \bibinfo{person}{Sagar Vemuri}, {and} \bibinfo{person}{David~H.C. Du}.}
  \bibinfo{year}{2020}\natexlab{}.
\newblock \showarticletitle{Characterizing, Modeling, and Benchmarking
  {RocksDB} Key-Value Workloads at {F}acebook}. In \bibinfo{booktitle}{{\em
  Proc. {USENIX} Conference on File and Storage Technologies ({FAST})}}.
\newblock
\showISBNx{978-1-939133-12-0}
\showURL{%
\url{https://www.usenix.org/conference/fast20/presentation/cao-zhichao}}


\bibitem[\protect\citeauthoryear{Chen, Ruan, Li, Ma, and Xu}{Chen
  et~al\mbox{.}}{2021}]%
        {spandb}
\bibfield{author}{\bibinfo{person}{Hao Chen}, \bibinfo{person}{Chaoyi Ruan},
  \bibinfo{person}{Cheng Li}, \bibinfo{person}{Xiaosong Ma}, {and}
  \bibinfo{person}{Yinlong Xu}.} \bibinfo{year}{2021}\natexlab{}.
\newblock \showarticletitle{Span{DB}: A Fast, Cost-Effective {LSM}-tree Based
  {KV} Store on Hybrid Storage}. In \bibinfo{booktitle}{{\em 19th {USENIX}
  Conference on File and Storage Technologies ({FAST} 21)}}.
  \bibinfo{pages}{17--32}.
\newblock


\bibitem[\protect\citeauthoryear{Cidon, Eisenman, Alizadeh, and Katti}{Cidon
  et~al\mbox{.}}{2016}]%
        {cliffhanger}
\bibfield{author}{\bibinfo{person}{Asaf Cidon}, \bibinfo{person}{Assaf
  Eisenman}, \bibinfo{person}{Mohammad Alizadeh}, {and} \bibinfo{person}{Sachin
  Katti}.} \bibinfo{year}{2016}\natexlab{}.
\newblock \showarticletitle{Cliffhanger: Scaling Performance Cliffs in Web
  Memory Caches}. In \bibinfo{booktitle}{{\em Proc. USENIX Symposium on
  Networked Systems Design and Implementation (NSDI)}}.
\newblock
\showISBNx{978-1-931971-29-4}
\showURL{%
\url{https://www.usenix.org/conference/nsdi16/technical-sessions/presentation/cidon}}


\bibitem[\protect\citeauthoryear{Cidon, Rushton, Rumble, and Stutsman}{Cidon
  et~al\mbox{.}}{2017}]%
        {memshare}
\bibfield{author}{\bibinfo{person}{Asaf Cidon}, \bibinfo{person}{Daniel
  Rushton}, \bibinfo{person}{Stephen~M. Rumble}, {and} \bibinfo{person}{Ryan
  Stutsman}.} \bibinfo{year}{2017}\natexlab{}.
\newblock \showarticletitle{Memshare: a Dynamic Multi-tenant Key-value Cache}.
  In \bibinfo{booktitle}{{\em 2017 {USENIX} Annual Technical Conference
  ({USENIX} {ATC} 17)}}. \bibinfo{publisher}{{USENIX} Association},
  \bibinfo{address}{Santa Clara, CA}, \bibinfo{pages}{321--334}.
\newblock
\showISBNx{978-1-931971-38-6}
\showURL{%
\url{https://www.usenix.org/conference/atc17/technical-sessions/presentation/cidon}}


\bibitem[\protect\citeauthoryear{Conway, Gupta, Chidambaram, Farach-Colton,
  Spillane, Tai, and Johnson}{Conway et~al\mbox{.}}{2020}]%
        {splinterdb}
\bibfield{author}{\bibinfo{person}{Alexander Conway}, \bibinfo{person}{Abhishek
  Gupta}, \bibinfo{person}{Vijay Chidambaram}, \bibinfo{person}{Martin
  Farach-Colton}, \bibinfo{person}{Richard Spillane}, \bibinfo{person}{Amy
  Tai}, {and} \bibinfo{person}{Rob Johnson}.} \bibinfo{year}{2020}\natexlab{}.
\newblock \showarticletitle{Splinter{DB}: Closing the Bandwidth Gap for {NVM}e
  Key-Value Stores}. In \bibinfo{booktitle}{{\em Proc. {USENIX} Annual
  Technical Conference (ATC)}}.
\newblock
\showISBNx{978-1-939133-14-4}
\showURL{%
\url{https://www.usenix.org/conference/atc20/presentation/conway}}


\bibitem[\protect\citeauthoryear{Cooper, Silberstein, Tam, Ramakrishnan, and
  Sears}{Cooper et~al\mbox{.}}{2010}]%
        {ycsb}
\bibfield{author}{\bibinfo{person}{Brian~F. Cooper}, \bibinfo{person}{Adam
  Silberstein}, \bibinfo{person}{Erwin Tam}, \bibinfo{person}{Raghu
  Ramakrishnan}, {and} \bibinfo{person}{Russell Sears}.}
  \bibinfo{year}{2010}\natexlab{}.
\newblock \showarticletitle{Benchmarking Cloud Serving Systems with {YCSB}}. In
  \bibinfo{booktitle}{{\em Proc. Symposium on Cloud Computing (SoCC)}}. 12.
\newblock
\showISBNx{9781450300360}
\showDOI{%
\url{https://doi.org/10.1145/1807128.1807152}}


\bibitem[\protect\citeauthoryear{Corbato}{Corbato}{1968}]%
        {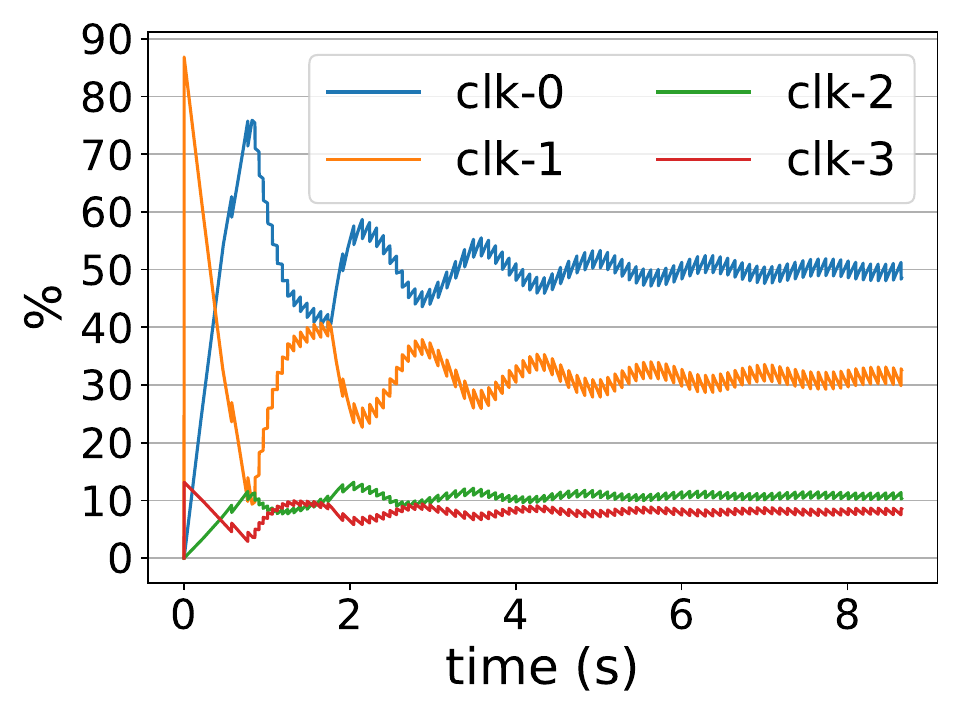}
\bibfield{author}{\bibinfo{person}{Fernando~J Corbato}.}
  \bibinfo{year}{1968}\natexlab{}.
\newblock \bibinfo{booktitle}{{\em A paging experiment with the multics
  system}}.
\newblock \bibinfo{type}{{T}echnical {R}eport}. \bibinfo{institution}{DTIC
  Document}.
\newblock


\bibitem[\protect\citeauthoryear{Corbett, Dean, Epstein, Fikes, Frost, Furman,
  Ghemawat, Gubarev, Heiser, Hochschild, Hsieh, Kanthak, Kogan, Li, Lloyd,
  Melnik, Mwaura, Nagle, Quinlan, Rao, Rolig, Saito, Szymaniak, Taylor, Wang,
  and Woodford}{Corbett et~al\mbox{.}}{2012}]%
        {spanner}
\bibfield{author}{\bibinfo{person}{James~C. Corbett}, \bibinfo{person}{Jeffrey
  Dean}, \bibinfo{person}{Michael Epstein}, \bibinfo{person}{Andrew Fikes},
  \bibinfo{person}{Christopher Frost}, \bibinfo{person}{JJ Furman},
  \bibinfo{person}{Sanjay Ghemawat}, \bibinfo{person}{Andrey Gubarev},
  \bibinfo{person}{Christopher Heiser}, \bibinfo{person}{Peter Hochschild},
  \bibinfo{person}{Wilson Hsieh}, \bibinfo{person}{Sebastian Kanthak},
  \bibinfo{person}{Eugene Kogan}, \bibinfo{person}{Hongyi Li},
  \bibinfo{person}{Alexander Lloyd}, \bibinfo{person}{Sergey Melnik},
  \bibinfo{person}{David Mwaura}, \bibinfo{person}{David Nagle},
  \bibinfo{person}{Sean Quinlan}, \bibinfo{person}{Rajesh Rao},
  \bibinfo{person}{Lindsay Rolig}, \bibinfo{person}{Yasushi Saito},
  \bibinfo{person}{Michal Szymaniak}, \bibinfo{person}{Christopher Taylor},
  \bibinfo{person}{Ruth Wang}, {and} \bibinfo{person}{Dale Woodford}.}
  \bibinfo{year}{2012}\natexlab{}.
\newblock \showarticletitle{Spanner: Google{\textquoteright}s
  Globally-Distributed Database}. In \bibinfo{booktitle}{{\em 10th USENIX
  Symposium on Operating Systems Design and Implementation (OSDI 12)}}.
  \bibinfo{address}{Hollywood, CA}, \bibinfo{pages}{261--264}.
\newblock
\showISBNx{978-1-931971-96-6}
\showURL{%
\url{https://www.usenix.org/conference/osdi12/technical-sessions/presentation/corbett}}


\bibitem[\protect\citeauthoryear{Dong, Callaghan, Galanis, Borthakur, Savor,
  and Strum}{Dong et~al\mbox{.}}{2017}]%
        {rocksdbcidr}
\bibfield{author}{\bibinfo{person}{Siying Dong}, \bibinfo{person}{Mark
  Callaghan}, \bibinfo{person}{Leonidas Galanis}, \bibinfo{person}{Dhruba
  Borthakur}, \bibinfo{person}{Tony Savor}, {and} \bibinfo{person}{Michael
  Strum}.} \bibinfo{year}{2017}\natexlab{}.
\newblock \showarticletitle{Optimizing Space Amplification in {R}ocks{DB}}. In
  \bibinfo{booktitle}{{\em Proc. Biennial Conference on Innovative Data Systems
  Research ({CIDR})}}.
\newblock
\showURL{%
\url{http://cidrdb.org/cidr2017/papers/p82-dong-cidr17.pdf}}


\bibitem[\protect\citeauthoryear{Eisenman, Cidon, Pergament, Haimovich,
  Stutsman, Alizadeh, and Katti}{Eisenman et~al\mbox{.}}{2019}]%
        {flashield}
\bibfield{author}{\bibinfo{person}{Assaf Eisenman}, \bibinfo{person}{Asaf
  Cidon}, \bibinfo{person}{Evgenya Pergament}, \bibinfo{person}{Or Haimovich},
  \bibinfo{person}{Ryan Stutsman}, \bibinfo{person}{Mohammad Alizadeh}, {and}
  \bibinfo{person}{Sachin Katti}.} \bibinfo{year}{2019}\natexlab{}.
\newblock \showarticletitle{Flashield: a Hybrid Key-value Cache that Controls
  Flash Write Amplification}. In \bibinfo{booktitle}{{\em Proc. {USENIX}
  Symposium on Networked Systems Design and Implementation ({NSDI})}}.
\newblock
\showISBNx{978-1-931971-49-2}
\showURL{%
\url{https://www.usenix.org/conference/nsdi19/presentation/eisenman}}


\bibitem[\protect\citeauthoryear{Eisenman, Gardner, AbdelRahman, Axboe, Dong,
  Hazelwood, Petersen, Cidon, and Katti}{Eisenman et~al\mbox{.}}{2018}]%
        {MyNVM}
\bibfield{author}{\bibinfo{person}{Assaf Eisenman}, \bibinfo{person}{Darryl
  Gardner}, \bibinfo{person}{Islam AbdelRahman}, \bibinfo{person}{Jens Axboe},
  \bibinfo{person}{Siying Dong}, \bibinfo{person}{Kim Hazelwood},
  \bibinfo{person}{Chris Petersen}, \bibinfo{person}{Asaf Cidon}, {and}
  \bibinfo{person}{Sachin Katti}.} \bibinfo{year}{2018}\natexlab{}.
\newblock \showarticletitle{Reducing {DRAM} Footprint with {NVM} in
  {F}acebook}. In \bibinfo{booktitle}{{\em Proc. EuroSys Conference}}. Article
  \bibinfo{articleno}{42}, \bibinfo{numpages}{13}~pages.
\newblock
\showISBNx{978-1-4503-5584-1}
\showDOI{%
\url{https://doi.org/10.1145/3190508.3190524}}


\bibitem[\protect\citeauthoryear{Fan, Andersen, and Kaminsky}{Fan
  et~al\mbox{.}}{2013}]%
        {fan2013memc3}
\bibfield{author}{\bibinfo{person}{Bin Fan}, \bibinfo{person}{David~G.
  Andersen}, {and} \bibinfo{person}{Michael Kaminsky}.}
  \bibinfo{year}{2013}\natexlab{}.
\newblock \showarticletitle{Mem{C3}: Compact and Concurrent {M}em{C}ache with
  Dumber Caching and Smarter Hashing}. In \bibinfo{booktitle}{{\em Proc. USENIX
  Symposium on Networked Systems Design and Implementation (NSDI)}}. 14.
\newblock
\showURL{%
\url{http://dl.acm.org/citation.cfm?id=2482626.2482662}}


\bibitem[\protect\citeauthoryear{Gilad, Bortnikov, Braginsky, Gottesman,
  Hillel, Keidar, Moscovici, and Shahout}{Gilad et~al\mbox{.}}{2020}]%
        {evendb}
\bibfield{author}{\bibinfo{person}{Eran Gilad}, \bibinfo{person}{Edward
  Bortnikov}, \bibinfo{person}{Anastasia Braginsky}, \bibinfo{person}{Yonatan
  Gottesman}, \bibinfo{person}{Eshcar Hillel}, \bibinfo{person}{Idit Keidar},
  \bibinfo{person}{Nurit Moscovici}, {and} \bibinfo{person}{Rana Shahout}.}
  \bibinfo{year}{2020}\natexlab{}.
\newblock \showarticletitle{Even{DB}: Optimizing Key-Value Storage for Spatial
  Locality}. In \bibinfo{booktitle}{{\em Proc. European Conference on Computer
  Systems (EuroSys)}}. Article \bibinfo{articleno}{27},
  \bibinfo{numpages}{16}~pages.
\newblock
\showISBNx{9781450368827}
\showDOI{%
\url{https://doi.org/10.1145/3342195.3387523}}


\bibitem[\protect\citeauthoryear{{Google b-tree implementation}}{{Google b-tree
  implementation}}{[n. d.]}]%
        {btree}
\bibfield{author}{\bibinfo{person}{{Google b-tree implementation}}.}
  \bibinfo{year}{[n. d.]}\natexlab{}.
\newblock \bibinfo{title}{https://code.google.com/archive/p/cpp-btree/}.
\newblock   (\bibinfo{year}{[n. d.]}).
\newblock
\showURL{%
\url{https://code.google.com/archive/p/cpp-btree/}}


\bibitem[\protect\citeauthoryear{Grupp, Davis, and Swanson}{Grupp
  et~al\mbox{.}}{2012}]%
        {bleak}
\bibfield{author}{\bibinfo{person}{Laura~M. Grupp}, \bibinfo{person}{John~D.
  Davis}, {and} \bibinfo{person}{Steven Swanson}.}
  \bibinfo{year}{2012}\natexlab{}.
\newblock \showarticletitle{The Bleak Future of {NAND} Flash Memory}. In
  \bibinfo{booktitle}{{\em Proc. USENIX Conference on File and Storage
  Technologies (FAST)}}. 1.
\newblock
\showURL{%
\url{http://dl.acm.org/citation.cfm?id=2208461.2208463}}


\bibitem[\protect\citeauthoryear{Intel}{Intel}{[n. d.]a}]%
        {p5800x}
\bibfield{author}{\bibinfo{person}{Intel}.} \bibinfo{year}{[n.
  d.]}\natexlab{a}.
\newblock \bibinfo{title}{{Intel Optane SSD DC P5800X Series}}.
\newblock   (\bibinfo{year}{[n. d.]}).
\newblock
\newblock
\shownote{\url{https://ark.intel.com/content/www/us/en/ark/products/201859/intel-optane-ssd-dc-p5800x-series-1-6tb-2-5in-pcie-x4-3d-xpoint.html}.}


\bibitem[\protect\citeauthoryear{Intel}{Intel}{[n. d.]b}]%
        {tbb}
\bibfield{author}{\bibinfo{person}{Intel}.} \bibinfo{year}{[n.
  d.]}\natexlab{b}.
\newblock \bibinfo{title}{Intel Thread Building Blocks (TBB) library}.
\newblock   (\bibinfo{year}{[n. d.]}).
\newblock
\newblock
\shownote{\url{https://software.intel.com/content/www/us/en/develop/tools/threading-building-blocks.html}.}


\bibitem[\protect\citeauthoryear{Jaffer, Mahdaviani, and Schroeder}{Jaffer
  et~al\mbox{.}}{2020}]%
        {jaffer2020rethinking}
\bibfield{author}{\bibinfo{person}{Shehbaz Jaffer}, \bibinfo{person}{Kaveh
  Mahdaviani}, {and} \bibinfo{person}{Bianca Schroeder}.}
  \bibinfo{year}{2020}\natexlab{}.
\newblock \showarticletitle{Rethinking {WOM} Codes to Enhance the Lifetime in
  New {SSD} Generations}. In \bibinfo{booktitle}{{\em Proc. {$USENIX$} Workshop
  on Hot Topics in Storage and File Systems (HotStorage)}}.
\newblock


\bibitem[\protect\citeauthoryear{Kaiyrakhmet, Lee, Nam, Noh, and
  Choi}{Kaiyrakhmet et~al\mbox{.}}{2019}]%
        {kaiyrakhmet2019slm}
\bibfield{author}{\bibinfo{person}{Olzhas Kaiyrakhmet}, \bibinfo{person}{Songyi
  Lee}, \bibinfo{person}{Beomseok Nam}, \bibinfo{person}{Sam~H Noh}, {and}
  \bibinfo{person}{Young-Ri Choi}.} \bibinfo{year}{2019}\natexlab{}.
\newblock \showarticletitle{{SLM-DB}: single-level key-value store with
  persistent memory}. In \bibinfo{booktitle}{{\em Proc. {USENIX} Conference on
  File and Storage Technologies (FAST)}}.
\newblock


\bibitem[\protect\citeauthoryear{Kourtis, Ioannou, and Koltsidas}{Kourtis
  et~al\mbox{.}}{2019}]%
        {udepot}
\bibfield{author}{\bibinfo{person}{Kornilios Kourtis}, \bibinfo{person}{Nikolas
  Ioannou}, {and} \bibinfo{person}{Ioannis Koltsidas}.}
  \bibinfo{year}{2019}\natexlab{}.
\newblock \showarticletitle{Reaping the performance of fast {NVM} storage with
  u{D}epot}. In \bibinfo{booktitle}{{\em Proc. {USENIX} Conference on File and
  Storage Technologies ({FAST})}}.
\newblock
\showISBNx{978-1-939133-09-0}
\showURL{%
\url{https://www.usenix.org/conference/fast19/presentation/kourtis}}


\bibitem[\protect\citeauthoryear{Kwon, Fingler, Hunt, Peter, Witchel, and
  Anderson}{Kwon et~al\mbox{.}}{2017}]%
        {kwon2017strata}
\bibfield{author}{\bibinfo{person}{Youngjin Kwon}, \bibinfo{person}{Henrique
  Fingler}, \bibinfo{person}{Tyler Hunt}, \bibinfo{person}{Simon Peter},
  \bibinfo{person}{Emmett Witchel}, {and} \bibinfo{person}{Thomas Anderson}.}
  \bibinfo{year}{2017}\natexlab{}.
\newblock \showarticletitle{Strata: A cross media file system}. In
  \bibinfo{booktitle}{{\em Proc. Symposium on Operating Systems Principles
  (SOSP)}}.
\newblock


\bibitem[\protect\citeauthoryear{Lepers, Balmau, Gupta, and Zwaenepoel}{Lepers
  et~al\mbox{.}}{2019}]%
        {KVell}
\bibfield{author}{\bibinfo{person}{Baptiste Lepers}, \bibinfo{person}{Oana
  Balmau}, \bibinfo{person}{Karan Gupta}, {and} \bibinfo{person}{Willy
  Zwaenepoel}.} \bibinfo{year}{2019}\natexlab{}.
\newblock \showarticletitle{{KVell}: The Design and Implementation of a Fast
  Persistent Key-Value Store}. In \bibinfo{booktitle}{{\em Proc. Symposium on
  Operating Systems Principles (SOSP)}}. 15.
\newblock
\showISBNx{9781450368735}
\showDOI{%
\url{https://doi.org/10.1145/3341301.3359628}}


\bibitem[\protect\citeauthoryear{Li, Jiang, Xiong, and Bao}{Li
  et~al\mbox{.}}{2020}]%
        {hilsm}
\bibfield{author}{\bibinfo{person}{Wenjie Li}, \bibinfo{person}{Dejun Jiang},
  \bibinfo{person}{Jin Xiong}, {and} \bibinfo{person}{Yungang Bao}.}
  \bibinfo{year}{2020}\natexlab{}.
\newblock \showarticletitle{{HiLSM: An LSM-Based Key-Value Store for Hybrid
  NVM-SSD Storage Systems}}. In \bibinfo{booktitle}{{\em Proceedings of the
  17th ACM International Conference on Computing Frontiers}} {\em
  (\bibinfo{series}{CF '20})}. \bibinfo{publisher}{Association for Computing
  Machinery}, \bibinfo{address}{New York, NY, USA}, \bibinfo{pages}{208–216}.
\newblock
\showISBNx{9781450379564}
\showDOI{%
\url{https://doi.org/10.1145/3387902.3392621}}


\bibitem[\protect\citeauthoryear{Marathe, Seltzer, Byan, and Harris}{Marathe
  et~al\mbox{.}}{2017}]%
        {persistentmemcached}
\bibfield{author}{\bibinfo{person}{Virendra~J. Marathe}, \bibinfo{person}{Margo
  Seltzer}, \bibinfo{person}{Steve Byan}, {and} \bibinfo{person}{Tim Harris}.}
  \bibinfo{year}{2017}\natexlab{}.
\newblock \showarticletitle{Persistent Memcached: Bringing Legacy Code to
  Byte-addressable Persistent Memory}. In \bibinfo{booktitle}{{\em Proc. USENIX
  Conference on Hot Topics in Storage and File Systems (HotStorage)}}. 1.
\newblock
\showURL{%
\url{http://dl.acm.org/citation.cfm?id=3154601.3154605}}


\bibitem[\protect\citeauthoryear{Mellor}{Mellor}{[n. d.]}]%
        {qlc}
\bibfield{author}{\bibinfo{person}{C Mellor}.} \bibinfo{year}{[n.
  d.]}\natexlab{}.
\newblock \bibinfo{title}{Toshiba flashes 100{TB} {QLC} flash drive, may go on
  sale within months. Really.}
\newblock   (\bibinfo{year}{[n. d.]}).
\newblock
\newblock
\shownote{\url{http://www.theregister.co.uk/2016/08/10/toshiba_100tb_qlc_ssd//}.}


\bibitem[\protect\citeauthoryear{Micron}{Micron}{[n. d.]a}]%
        {micronwhitepaper}
\bibfield{author}{\bibinfo{person}{Micron}.} \bibinfo{year}{[n.
  d.]}\natexlab{a}.
\newblock \bibinfo{title}{Comparing {SSD} and {HDD} Endurance in the Age of
  {QLC SSD}s}.
\newblock   (\bibinfo{year}{[n. d.]}).
\newblock
\newblock
\shownote{\url{https://www.micron.com/-/media/client/global/documents/products/white-paper/5210_ssd_vs_hdd_endurance_white_paper.pdf}.}


\bibitem[\protect\citeauthoryear{Micron}{Micron}{[n. d.]b}]%
        {micronqlc}
\bibfield{author}{\bibinfo{person}{Micron}.} \bibinfo{year}{[n.
  d.]}\natexlab{b}.
\newblock \bibinfo{title}{{QLC NAND} Technology}.
\newblock   (\bibinfo{year}{[n. d.]}).
\newblock
\newblock
\shownote{\url{https://www.micron.com/products/advanced-solutions/qlc-nand}.}


\bibitem[\protect\citeauthoryear{Mitzenmacher}{Mitzenmacher}{2001}]%
        {power-of-two}
\bibfield{author}{\bibinfo{person}{Michael Mitzenmacher}.}
  \bibinfo{year}{2001}\natexlab{}.
\newblock \showarticletitle{The power of two choices in randomized load
  balancing}.
\newblock \bibinfo{journal}{{\em IEEE Transactions on Parallel and Distributed
  Systems\/}} \bibinfo{volume}{12}, \bibinfo{number}{10}
  (\bibinfo{year}{2001}), \bibinfo{pages}{1094--1104}.
\newblock


\bibitem[\protect\citeauthoryear{Nishtala, Fugal, Grimm, Kwiatkowski, Lee, Li,
  McElroy, Paleczny, Peek, Saab, Stafford, Tung, and Venkataramani}{Nishtala
  et~al\mbox{.}}{2013}]%
        {nishtala2013scaling}
\bibfield{author}{\bibinfo{person}{Rajesh Nishtala}, \bibinfo{person}{Hans
  Fugal}, \bibinfo{person}{Steven Grimm}, \bibinfo{person}{Marc Kwiatkowski},
  \bibinfo{person}{Herman Lee}, \bibinfo{person}{Harry~C. Li},
  \bibinfo{person}{Ryan McElroy}, \bibinfo{person}{Mike Paleczny},
  \bibinfo{person}{Daniel Peek}, \bibinfo{person}{Paul Saab},
  \bibinfo{person}{David Stafford}, \bibinfo{person}{Tony Tung}, {and}
  \bibinfo{person}{Venkateshwaran Venkataramani}.}
  \bibinfo{year}{2013}\natexlab{}.
\newblock \showarticletitle{Scaling {M}emcache at {F}acebook}. In
  \bibinfo{booktitle}{{\em Proc. USENIX Symposium on Networked Systems Design
  and Implementation (NSDI)}}.
\newblock
\showISBNx{978-1-931971-00-3}


\bibitem[\protect\citeauthoryear{Ogleari, Miller, and Zhao}{Ogleari
  et~al\mbox{.}}{2018}]%
        {ogleari2018steal}
\bibfield{author}{\bibinfo{person}{Matheus~Almeida Ogleari},
  \bibinfo{person}{Ethan~L Miller}, {and} \bibinfo{person}{Jishen Zhao}.}
  \bibinfo{year}{2018}\natexlab{}.
\newblock \showarticletitle{Steal but no force: Efficient hardware undo+ redo
  logging for persistent memory systems}. In \bibinfo{booktitle}{{\em Proc.
  Intl. Symposium on High Performance Computer Architecture (HPCA)}}.
\newblock


\bibitem[\protect\citeauthoryear{Ohshima and Tanaka}{Ohshima and Tanaka}{[n.
  d.]}]%
        {flashmemorysummit}
\bibfield{author}{\bibinfo{person}{S Ohshima} {and} \bibinfo{person}{Y
  Tanaka}.} \bibinfo{year}{[n. d.]}\natexlab{}.
\newblock \bibinfo{title}{New {3D} Flash Technologies Offer Both Low Cost and
  Low Power Solutions}.
\newblock   (\bibinfo{year}{[n. d.]}).
\newblock
\newblock
\shownote{\url{https://www.flashmemorysummit.com/English/Conference/Keynotes.html}.}


\bibitem[\protect\citeauthoryear{O'Neil, Cheng, Gawlick, and O'Neil}{O'Neil
  et~al\mbox{.}}{1996}]%
        {o1996log}
\bibfield{author}{\bibinfo{person}{Patrick O'Neil}, \bibinfo{person}{Edward
  Cheng}, \bibinfo{person}{Dieter Gawlick}, {and} \bibinfo{person}{Elizabeth
  O'Neil}.} \bibinfo{year}{1996}\natexlab{}.
\newblock \showarticletitle{The log-structured merge-tree ({LSM}-tree)}.
\newblock \bibinfo{journal}{{\em Acta Informatica\/}} \bibinfo{volume}{33},
  \bibinfo{number}{4} (\bibinfo{year}{1996}).
\newblock


\bibitem[\protect\citeauthoryear{Raju, Kadekodi, Chidambaram, and Abraham}{Raju
  et~al\mbox{.}}{2017}]%
        {pebblesdb}
\bibfield{author}{\bibinfo{person}{Pandian Raju}, \bibinfo{person}{Rohan
  Kadekodi}, \bibinfo{person}{Vijay Chidambaram}, {and} \bibinfo{person}{Ittai
  Abraham}.} \bibinfo{year}{2017}\natexlab{}.
\newblock \showarticletitle{Pebbles{DB}: Building Key-Value Stores Using
  Fragmented Log-Structured Merge Trees}. In \bibinfo{booktitle}{{\em Proc.
  Symposium on Operating Systems Principles (SOSP)}}. 18.
\newblock
\showISBNx{978-1-4503-5085-3}
\showDOI{%
\url{https://doi.org/10.1145/3132747.3132765}}


\bibitem[\protect\citeauthoryear{{RocksDB}}{{RocksDB}}{[n. d.]}]%
        {rocksdb}
\bibfield{author}{\bibinfo{person}{{RocksDB}}.} \bibinfo{year}{[n.
  d.]}\natexlab{}.
\newblock \bibinfo{title}{https://rocksdb.org}.
\newblock   (\bibinfo{year}{[n. d.]}).
\newblock
\showURL{%
\url{https://rocksdb.org}}


\bibitem[\protect\citeauthoryear{Rosenblum and Ousterhout}{Rosenblum and
  Ousterhout}{1992}]%
        {LFS}
\bibfield{author}{\bibinfo{person}{Mendel Rosenblum} {and}
  \bibinfo{person}{John~K Ousterhout}.} \bibinfo{year}{1992}\natexlab{}.
\newblock \showarticletitle{The design and implementation of a log-structured
  file system}.
\newblock \bibinfo{journal}{{\em ACM Trans. Computer Systems\/}}
  \bibinfo{volume}{10}, \bibinfo{number}{1} (\bibinfo{year}{1992}).
\newblock


\bibitem[\protect\citeauthoryear{Rumble, Kejriwal, and Ousterhout}{Rumble
  et~al\mbox{.}}{2014}]%
        {lsm-ramcloud}
\bibfield{author}{\bibinfo{person}{Stephen~M. Rumble}, \bibinfo{person}{Ankita
  Kejriwal}, {and} \bibinfo{person}{John Ousterhout}.}
  \bibinfo{year}{2014}\natexlab{}.
\newblock \showarticletitle{Log-structured Memory for {DRAM-based} Storage}. In
  \bibinfo{booktitle}{{\em 12th USENIX Conference on File and Storage
  Technologies (FAST 14)}}. \bibinfo{publisher}{USENIX Association},
  \bibinfo{address}{Santa Clara, CA}, \bibinfo{pages}{1--16}.
\newblock
\showISBNx{ISBN 978-1-931971-08-9}
\showURL{%
\url{https://www.usenix.org/conference/fast14/technical-sessions/presentation/rumble}}


\bibitem[\protect\citeauthoryear{Song, Berger, Li, and Lloyd}{Song
  et~al\mbox{.}}{2020}]%
        {lrb}
\bibfield{author}{\bibinfo{person}{Zhenyu Song}, \bibinfo{person}{Daniel~S.
  Berger}, \bibinfo{person}{Kai Li}, {and} \bibinfo{person}{Wyatt Lloyd}.}
  \bibinfo{year}{2020}\natexlab{}.
\newblock \showarticletitle{Learning Relaxed {B}elady for Content Distribution
  Network Caching}. In \bibinfo{booktitle}{{\em Proc. {USENIX} Symposium on
  Networked Systems Design and Implementation ({NSDI})}}.
\newblock
\showISBNx{978-1-939133-13-7}
\showURL{%
\url{https://www.usenix.org/conference/nsdi20/presentation/song}}


\bibitem[\protect\citeauthoryear{Tai, Kryczka, Kanaujia, Jamieson, Freedman,
  and Cidon}{Tai et~al\mbox{.}}{2019}]%
        {DIRECT}
\bibfield{author}{\bibinfo{person}{Amy Tai}, \bibinfo{person}{Andrew Kryczka},
  \bibinfo{person}{Shobhit~O. Kanaujia}, \bibinfo{person}{Kyle Jamieson},
  \bibinfo{person}{Michael~J. Freedman}, {and} \bibinfo{person}{Asaf Cidon}.}
  \bibinfo{year}{2019}\natexlab{}.
\newblock \showarticletitle{Who{\textquoteright}s Afraid of Uncorrectable Bit
  Errors? {O}nline Recovery of Flash Errors with Distributed Redundancy}. In
  \bibinfo{booktitle}{{\em Proc. {USENIX} Annual Technical Conference (ATC)}}.
\newblock
\showISBNx{978-1-939133-03-8}
\showURL{%
\url{https://www.usenix.org/conference/atc19/presentation/tai}}


\bibitem[\protect\citeauthoryear{Tallis}{Tallis}{[n. d.]}]%
        {lowpe}
\bibfield{author}{\bibinfo{person}{Billy Tallis}.} \bibinfo{year}{[n.
  d.]}\natexlab{}.
\newblock \bibinfo{title}{The Crucial {P}1 1{TB SSD} Review: The Other Consumer
  {QLC SSD}}.
\newblock   (\bibinfo{year}{[n. d.]}).
\newblock
\newblock
\shownote{\url{https://www.anandtech.com/show/13512/the-crucial-p1-1tb-ssd-review}.}


\bibitem[\protect\citeauthoryear{Tang, Huang, Lloyd, Kumar, and Li}{Tang
  et~al\mbox{.}}{2015}]%
        {RIPQ}
\bibfield{author}{\bibinfo{person}{Linpeng Tang}, \bibinfo{person}{Qi Huang},
  \bibinfo{person}{Wyatt Lloyd}, \bibinfo{person}{Sanjeev Kumar}, {and}
  \bibinfo{person}{Kai Li}.} \bibinfo{year}{2015}\natexlab{}.
\newblock \showarticletitle{{RIPQ}: Advanced Photo Caching on Flash for
  {F}acebook}. In \bibinfo{booktitle}{{\em Proc. USENIX Conference on File and
  Storage Technologies (FAST)}}.
\newblock
\showISBNx{978-1-931971-201}
\showURL{%
\url{https://www.usenix.org/conference/fast15/technical-sessions/presentation/tang}}


\bibitem[\protect\citeauthoryear{{Tracking live sst files}}{{Tracking live sst
  files}}{[n. d.]}]%
        {versionset}
\bibfield{author}{\bibinfo{person}{{Tracking live sst files}}.}
  \bibinfo{year}{[n. d.]}\natexlab{}.
\newblock
  \bibinfo{title}{https://github.com/facebook/rocksdb/wiki/How-we-keep-track-of-live-SST-files}.
\newblock   (\bibinfo{year}{[n. d.]}).
\newblock
\showURL{%
\url{https://github.com/facebook/rocksdb/wiki/How-we-keep-track-of-live-SST-files}}


\bibitem[\protect\citeauthoryear{van Renen, Leis, Kemper, Neumann, Hashida, Oe,
  Doi, Harada, and Sato}{van Renen et~al\mbox{.}}{2018}]%
        {vanrenen}
\bibfield{author}{\bibinfo{person}{Alexander van Renen},
  \bibinfo{person}{Viktor Leis}, \bibinfo{person}{Alfons Kemper},
  \bibinfo{person}{Thomas Neumann}, \bibinfo{person}{Takushi Hashida},
  \bibinfo{person}{Kazuichi Oe}, \bibinfo{person}{Yoshiyasu Doi},
  \bibinfo{person}{Lilian Harada}, {and} \bibinfo{person}{Mitsuru Sato}.}
  \bibinfo{year}{2018}\natexlab{}.
\newblock \showarticletitle{Managing Non-Volatile Memory in Database Systems}.
  In \bibinfo{booktitle}{{\em Proc. Intl. Conference on Management of Data
  (SIGMOD)}}. 15.
\newblock
\showISBNx{9781450347037}
\showDOI{%
\url{https://doi.org/10.1145/3183713.3196897}}


\bibitem[\protect\citeauthoryear{van Renen, Vogel, Leis, Neumann, and
  Kemper}{van Renen et~al\mbox{.}}{2019}]%
        {vanrenen2}
\bibfield{author}{\bibinfo{person}{Alexander van Renen}, \bibinfo{person}{Lukas
  Vogel}, \bibinfo{person}{Viktor Leis}, \bibinfo{person}{Thomas Neumann},
  {and} \bibinfo{person}{Alfons Kemper}.} \bibinfo{year}{2019}\natexlab{}.
\newblock \showarticletitle{Persistent Memory {I/O} Primitives}. In
  \bibinfo{booktitle}{{\em Proc. Intl. Workshop on Data Management on New
  Hardware (DaMoN)}}. Article \bibinfo{articleno}{Article 12},
  \bibinfo{numpages}{7}~pages.
\newblock
\showISBNx{9781450368018}
\showDOI{%
\url{https://doi.org/10.1145/3329785.3329930}}


\bibitem[\protect\citeauthoryear{Waldspurger, Saemundsson, Ahmad, and
  Park}{Waldspurger et~al\mbox{.}}{2017}]%
        {mini-caches}
\bibfield{author}{\bibinfo{person}{Carl Waldspurger}, \bibinfo{person}{Trausti
  Saemundsson}, \bibinfo{person}{Irfan Ahmad}, {and} \bibinfo{person}{Nohhyun
  Park}.} \bibinfo{year}{2017}\natexlab{}.
\newblock \showarticletitle{Cache Modeling and Optimization using Miniature
  Simulations}. In \bibinfo{booktitle}{{\em Proc. {USENIX} Annual Technical
  Conference (ATC)}}.
\newblock
\showISBNx{978-1-931971-38-6}
\showURL{%
\url{https://www.usenix.org/conference/atc17/technical-sessions/presentation/waldspurger}}


\bibitem[\protect\citeauthoryear{Waldspurger, Park, Garthwaite, and
  Ahmad}{Waldspurger et~al\mbox{.}}{2015}]%
        {SHARDS}
\bibfield{author}{\bibinfo{person}{Carl~A. Waldspurger},
  \bibinfo{person}{Nohhyun Park}, \bibinfo{person}{Alexander Garthwaite}, {and}
  \bibinfo{person}{Irfan Ahmad}.} \bibinfo{year}{2015}\natexlab{}.
\newblock \showarticletitle{Efficient {MRC} Construction with {SHARDS}}. In
  \bibinfo{booktitle}{{\em Proc. {USENIX} Conference on File and Storage
  Technologies ({FAST})}}.
\newblock
\showISBNx{978-1-931971-201}
\showURL{%
\url{https://www.usenix.org/conference/fast15/technical-sessions/presentation/waldspurger}}


\bibitem[\protect\citeauthoryear{Walker}{Walker}{2016}]%
        {spdk}
\bibfield{author}{\bibinfo{person}{Benjamin Walker}.}
  \bibinfo{year}{2016}\natexlab{}.
\newblock \showarticletitle{{SPDK}: Building blocks for scalable, high
  performance storage applications}. In \bibinfo{booktitle}{{\em Storage
  Developer Conference. SNIA}}.
\newblock


\bibitem[\protect\citeauthoryear{Webster}{Webster}{[n. d.]}]%
        {660p}
\bibfield{author}{\bibinfo{person}{Sean Webster}.} \bibinfo{year}{[n.
  d.]}\natexlab{}.
\newblock \bibinfo{title}{Intel {SSD} 660p}.
\newblock   (\bibinfo{year}{[n. d.]}).
\newblock
\newblock
\shownote{\url{https://www.tomshardware.com/reviews/intel-ssd-660p-qlc-nvme,5719.html}.}


\bibitem[\protect\citeauthoryear{Wu, Arpaci-Dusseau, and Arpaci-Dusseau}{Wu
  et~al\mbox{.}}{2019a}]%
        {unwritten-contract}
\bibfield{author}{\bibinfo{person}{Kan Wu}, \bibinfo{person}{Andrea
  Arpaci-Dusseau}, {and} \bibinfo{person}{Remzi Arpaci-Dusseau}.}
  \bibinfo{year}{2019}\natexlab{a}.
\newblock \showarticletitle{Towards an Unwritten Contract of {Intel Optane}
  {SSD}}. In \bibinfo{booktitle}{{\em Proc. {USENIX} Workshop on Hot Topics in
  Storage and File Systems (HotStorage)}}.
\newblock
\showURL{%
\url{https://www.usenix.org/conference/hotstorage19/presentation/wu-kan}}


\bibitem[\protect\citeauthoryear{Wu, Arpaci-Dusseau, Arpaci-Dusseau, Sen, and
  Park}{Wu et~al\mbox{.}}{2019b}]%
        {exploitingoptane}
\bibfield{author}{\bibinfo{person}{Kan Wu}, \bibinfo{person}{Andrea
  Arpaci-Dusseau}, \bibinfo{person}{Remzi Arpaci-Dusseau},
  \bibinfo{person}{Rathijit Sen}, {and} \bibinfo{person}{Kwanghyun Park}.}
  \bibinfo{year}{2019}\natexlab{b}.
\newblock \showarticletitle{{Exploiting Intel Optane SSD for Microsoft SQL
  Server}}. In \bibinfo{booktitle}{{\em Proc. Intl. Workshop on Data Management
  on New Hardware (DaMoN)}}. Article \bibinfo{articleno}{15},
  \bibinfo{numpages}{3}~pages.
\newblock
\showISBNx{9781450368018}
\showDOI{%
\url{https://doi.org/10.1145/3329785.3329916}}


\bibitem[\protect\citeauthoryear{Wu, Guo, Hu, Tu, Alagappan, Sen, Park,
  Arpaci-Dusseau, and Arpaci-Dusseau}{Wu et~al\mbox{.}}{2021}]%
        {orthus}
\bibfield{author}{\bibinfo{person}{Kan Wu}, \bibinfo{person}{Zhihan Guo},
  \bibinfo{person}{Guanzhou Hu}, \bibinfo{person}{Kaiwei Tu},
  \bibinfo{person}{Ramnatthan Alagappan}, \bibinfo{person}{Rathijit Sen},
  \bibinfo{person}{Kwanghyun Park}, \bibinfo{person}{Andrea~C. Arpaci-Dusseau},
  {and} \bibinfo{person}{Remzi~H. Arpaci-Dusseau}.}
  \bibinfo{year}{2021}\natexlab{}.
\newblock \showarticletitle{The Storage Hierarchy is Not a Hierarchy:
  Optimizing Caching on Modern Storage Devices with {O}rthus}. In
  \bibinfo{booktitle}{{\em 19th {USENIX} Conference on File and Storage
  Technologies ({FAST} 21)}}. \bibinfo{publisher}{{USENIX} Association},
  \bibinfo{pages}{307--323}.
\newblock
\showISBNx{978-1-939133-20-5}
\showURL{%
\url{https://www.usenix.org/conference/fast21/presentation/wu-kan}}


\bibitem[\protect\citeauthoryear{Xia, Jiang, Xiong, and Sun}{Xia
  et~al\mbox{.}}{2017}]%
        {HiKV}
\bibfield{author}{\bibinfo{person}{Fei Xia}, \bibinfo{person}{Dejun Jiang},
  \bibinfo{person}{Jin Xiong}, {and} \bibinfo{person}{Ninghui Sun}.}
  \bibinfo{year}{2017}\natexlab{}.
\newblock \showarticletitle{Hi{KV}: A Hybrid Index Key-Value Store for
  {DRAM-NVM} Memory Systems}. In \bibinfo{booktitle}{{\em Proc. {USENIX} Annual
  Technical Conference ({ATC})}}.
\newblock
\showISBNx{978-1-931971-38-6}
\showURL{%
\url{https://www.usenix.org/conference/atc17/technical-sessions/presentation/xia}}


\bibitem[\protect\citeauthoryear{Yang, Yue, and Rashmi}{Yang
  et~al\mbox{.}}{2020}]%
        {twitter-cache}
\bibfield{author}{\bibinfo{person}{Juncheng Yang}, \bibinfo{person}{Yao Yue},
  {and} \bibinfo{person}{K.~V. Rashmi}.} \bibinfo{year}{2020}\natexlab{}.
\newblock \showarticletitle{A large scale analysis of hundreds of in-memory
  cache clusters at {T}witter}. In \bibinfo{booktitle}{{\em 14th {USENIX}
  Symposium on Operating Systems Design and Implementation ({OSDI} 20)}}.
  \bibinfo{publisher}{{USENIX} Association}, \bibinfo{pages}{191--208}.
\newblock
\showISBNx{978-1-939133-19-9}
\showURL{%
\url{https://www.usenix.org/conference/osdi20/presentation/yang}}


\bibitem[\protect\citeauthoryear{Yao, Zhang, Wan, Cui, Tang, Jiang, Xie, and
  He}{Yao et~al\mbox{.}}{2020}]%
        {matrixkv}
\bibfield{author}{\bibinfo{person}{Ting Yao}, \bibinfo{person}{Yiwen Zhang},
  \bibinfo{person}{Jiguang Wan}, \bibinfo{person}{Qiu Cui},
  \bibinfo{person}{Liu Tang}, \bibinfo{person}{Hong Jiang},
  \bibinfo{person}{Changsheng Xie}, {and} \bibinfo{person}{Xubin He}.}
  \bibinfo{year}{2020}\natexlab{}.
\newblock \showarticletitle{Matrix{KV}: Reducing Write Stalls and Write
  Amplification in LSM-tree Based {KV} Stores with Matrix Container in {NVM}}.
  In \bibinfo{booktitle}{{\em Proc. {USENIX} Annual Technical Conference
  (ATC)}}.
\newblock
\showISBNx{978-1-939133-14-4}
\showURL{%
\url{https://www.usenix.org/conference/atc20/presentation/yao}}


\bibitem[\protect\citeauthoryear{Yoon, Yang, Kristjansson, Sigurdarson,
  Vigfusson, and Gavrilovska}{Yoon et~al\mbox{.}}{2018}]%
        {mutant}
\bibfield{author}{\bibinfo{person}{Hobin Yoon}, \bibinfo{person}{Juncheng
  Yang}, \bibinfo{person}{Sveinn~Fannar Kristjansson},
  \bibinfo{person}{Steinn~E. Sigurdarson}, \bibinfo{person}{Ymir Vigfusson},
  {and} \bibinfo{person}{Ada Gavrilovska}.} \bibinfo{year}{2018}\natexlab{}.
\newblock \showarticletitle{Mutant: Balancing Storage Cost and Latency in
  {LSM}-Tree Data Stores}. In \bibinfo{booktitle}{{\em Proc. Symposium on Cloud
  Computing (SoCC)}}. 12.
\newblock
\showISBNx{9781450360111}
\showDOI{%
\url{https://doi.org/10.1145/3267809.3267846}}


\bibitem[\protect\citeauthoryear{Zhou, Arulraj, Pavlo, and Cohen}{Zhou
  et~al\mbox{.}}{2021}]%
        {spitfire}
\bibfield{author}{\bibinfo{person}{Xinjing Zhou}, \bibinfo{person}{Joy
  Arulraj}, \bibinfo{person}{Andrew Pavlo}, {and} \bibinfo{person}{David
  Cohen}.} \bibinfo{year}{2021}\natexlab{}.
\newblock \bibinfo{booktitle}{{\em Spitfire: A Three-Tier Buffer Manager for
  Volatile and Non-Volatile Memory}}.
\newblock \bibinfo{publisher}{Association for Computing Machinery},
  \bibinfo{address}{New York, NY, USA}, \bibinfo{pages}{2195–2207}.
\newblock
\showISBNx{9781450383431}
\showURL{%
\url{https://doi.org/10.1145/3448016.3452819}}


\end{thebibliography}

\end{document}